\documentclass{ws-ijmpd}
\usepackage{slashed}
\usepackage{bbold}
\usepackage{graphicx}
\newcommand{\Tr}{\ensuremath{\text{Tr}}}

\newcommand{\be}{\begin{equation}}
\newcommand{\ee}{\end{equation}}
\newcommand{\bea}{\begin{eqnarray}}
\newcommand{\eea}{\end{eqnarray}}
\newcommand{\br}{\hskip .25cm/\hskip -.25cm}

\newcommand{\nn}{\nonumber\\}

\begin{document}

\markboth{J. Alexandre, N. Houston and N.E. Mavromatos}
{Inflation and Dynamically Broken Supergravity}


\title{INFLATION VIA GRAVITINO CONDENSATION IN DYNAMICALLY BROKEN SUPERGRAVITY}

\author{JEAN ALEXANDRE, NICK HOUSTON and NICK E. MAVROMATOS\footnote{Also currently at:  Theory Division, Physics Department, CERN, CH-1211 Geneva 23, Switzerland.}\, \footnote{Corresponding author.}}

\address{Theoretical Particle Physics and Cosmology Group, Physics Department, King's College London, Strand, London WC2R 2LS, UK. }

\maketitle

%


\begin{abstract} 
Gravitino-condensate-induced inflation via the super-Higgs effect is a UV-motivated scenario for both inflating the early universe and breaking local supersymmetry dynamically, entirely independent of any coupling to external matter.
As an added benefit, this also removes the (as of yet unobserved) massless Goldstino associated to global supersymmetry breaking from the particle spectrum. In this review
we detail the pertinent properties and outline previously hidden details of the various steps required in this context in order to make contact with current inflationary phenomenology. 
The class of models of SUGRA we use to exemplify our approach are minimal four-dimensional N=1 supergravity and conformal extensions thereof (with broken conformal symmetry). Therein, the gravitino condensate itself can play the role of the inflaton, however the requirement of slow-roll necessitates unnaturally large values of the wave-function renormalisation. Nevertheless, there is 
an alternative scenario that may provide Starobinsky-type inflation, occurring in the broken-SUGRA phase around the non-trivial minima of the gravitino-condensate effective potential. In this scenario higher curvature corrections to the effective action, crucial for the onset of an inflationary phase, arise as a result of integrating out massive quantum gravitino fields in the path integral. The latter scenario is compatible with Planck satellite phenomenology but not with BICEP2 data.

\end{abstract}


\ccode{PACS numbers:\, 98.80.-k;04.65.+e;11.15.Ex;11.30.Qc}

\vspace{0.2cm} 
KCL-PH-TH/2014-{\bf 20},  LCTS/2014-20  \emph{Invited Review}, special issue IJMPD

\section{Introduction}

The inflationary paradigm is at present a successful one, offering an elegant solution to the so-called horizon and flatness problems of the standard Big Bang cosmology, whilst simultaneously seeding both the large-scale structure of the universe and temperature anisotropies of the CMB via quantum fluctuations occurring during the inflationary epoch.
The precise microphysical mechanism of inflation is however unknown at present.

The data favour, or - from a rather more conservative viewpoint - are in agreement with, a scalar field or fields with canonical kinetic terms slowly rolling down an almost flat potential in the context of Einstein gravity, generating in the process 50 - 60 e-folds of inflation, along with adiabatic, nearly scale invariant primordial density perturbations \cite{Planck,encyclo}.

An important issue at present is the extent to which this inflationary process is tied to physics at the Grand Unification  (GUT) scale, and in particular, to a possible supersymmetric phase transition occurring in the early universe. 
Links of supersymmetry to inflation may be arguably expected from the fact that supersymmetry provides a rather natural reason~\cite{natural} for the observational fact~\cite{Planck} that the Hubble scale of inflation is much smaller than 
the Planck scale, lying in the ballpark of the GUT scale
\be\label{HI}
H_I \leq  0.74 \times 10^{-5} \, m_P = {\mathcal O}(10^{15})~{\rm GeV}~.
\ee
If supersymmetry is realised in nature however, it is certainly broken. 

It is known that simple realisations of global supersymmetry (SUSY) breaking, such as in the Wess-Zumino model~\cite{croon}, can provide, 
when embedded in gravitational environments, slow-roll models for inflation consistent with both Planck~\cite{Planck} and BICEP2~\cite{BICEP2} data.
Rigorous embeddings of global SUSY to local supersymmetry (SUGRA) have also been considered and explored in the literature over the years in connection with various scenarios for inflation, such as hybrid~\cite{sugra_hybrid}, chaotic~\cite{sugra_chaotic}, no-scale SUGRA/Starobinsky-like~\cite{sugra_staro}. 
In the latter case inflation is linked to higher curvature terms in the gravitational action (such as $R^2$ terms), as in the original
Starobinsky model~\cite{staro}, and others~\cite{confsugra,Ferrara:2010in,sugrainfl}.

For a recent review on supergravity and inflation we refer the reader to  ref.~\refcite{sugra_inflation_review}. Such models have been compared against the recently available data, with the conclusion that, although Planck data~\cite{Planck} compatibility is straightforward, the surprisingly large ratio of tensor-to-scalar primordial fluctuations, 
\be\label{large}
r=0.16_{-0.05}^{+0.06} \quad  ({\rm after~foreground~subtraction})~,
\ee
claimed to have been observed by the BICEP2~\cite{BICEP2} collaboration, presents in general a challenge. 
Needless to say there is a tension present between the BICEP2 and Planck results, with Planck favouring $r<0.11$ at the 95\% confidence level \cite{Planck}. Indeed, 
From the best fit value of the running spectral index $n_s\sim0.96$ found by Planck \cite{Planck}, which BICEP2 agrees with, and the usual relations among the slow-roll inflationary parameters~\cite{encyclo}
\begin{align}\label{spectral}
	n_s=1-6\epsilon+2\eta\,, \quad 
	r=16\epsilon\,,
\end{align}
we then find $r\lesssim0.11$.
The BICEP2 measurement still needs to be confirmed by Planck and other future experiments. 

In some previous publications~\cite{emdyno,ahm,ahmstaro} we have discussed the possibility of dynamically breaking SUGRA solely by means of exploiting the four-gravitino interactions that characterise (any) supergravity action, via the fermionic torsion parts of the spin connection. 
The primary example, where the calculations of the effective potential were detailed, was that of $\mathcal{N}=1$, $D=4$ simple SUGRA without matter~\cite{Freedman,Nieuwenhuizen}. 

The dynamical breaking process may be concretely realised by means of a phase transition from the supersymmetric phase where the bilinear $\langle\overline\psi_\mu\psi^\mu\rangle$ representing the effective scalar degree of freedom has zero vacuum expectation value, to one where $\sigma \equiv \langle\overline\psi_\mu\psi^\mu\rangle\neq0$. 
The quantum excitations about this condensate vacuum are then identified with a gravitino condensate field. 
Since this must be an energetically favourable process to occur, it then follows that the effective potential experienced by the gravitino condensate must be locally concave about the origin. 

The corresponding one-loop effective potential of the gravitino condensate field, obtained after integrating out fermionic (gravitino) and bosonic (graviton) degrees of freedom therefore has the characteristic form of a Coleman-Weinberg double well potential, offering the possibility of hilltop-type inflation, with the condensate field playing the role of the inflaton~\cite{emdyno,ahm}$^,$\footnote{We note at this stage that our gravitino-condensate model of inflation is rather different from the model of minimal inflation of ref.~\refcite{alvarez}. 
There, inflation is realised in the Ultraviolet via the scalar component $x$ of the so-called Ferrara-Zumino current superfield, $X$,  which in the Infrared becomes a two Goldstino state~\cite{Komargodski}, since the superfield satisfies a non-linear constraint $X^2 = 0$.  
An F-type supersymmetry-breaking effective superpotential for $X$ was assumed in ref.~\refcite{alvarez}, and a potential for $x$ was induced from gravitational corrections to the appropriate Kahler potential. 
In the present scenario however we only deal with gravitino condensate fields, whose one loop effective potential is obtained, as already mentioned, by integrating out gravitino and graviton degrees of freedom. 
The explicit form of the superpotential responsible for global supersymmetry breaking, as well as the associated Kahler potential for SUGRA, are not relevant for our minimal scenario for inflation.}. However, Starobinsky-type inflation~\cite{staro} may also be a possibility in the massive gravitino phase, as a result of the conformal anomaly induced by the development of a gravitino mass~\cite{ahmstaro}. 

There are a number of advantages to the gravitino-condensate scenario for inflation:
\begin{itemize}
	\item Principally, the formation of this condensate may both inflate the early universe and break local supersymmetry simultaneously, requiring the gravitino field to perform `double duty' and forcing the model to confront both inflationary and particle physics phenomenology.
	\item This process may occur independently of coupling to external matter; in contrast to other supersymmetry breaking scenarios, offering a certain universality within the context of supergravity.
	\item From an ultraviolet perspective it is attractive to realise inflation within the context of supergravity theories, which are thought to constitute consistent low energy limits of string/M theories.
	\item By virtue of the super-Higgs effect, the gravitino `eats' the (as of yet unobserved) massless goldstino associated to global supersymmetry breaking, thus removing it from the particle spectrum \cite{DeserZumino}.
\end{itemize}

\begin{figure}[h!!]
  \centering
    	\includegraphics[width=0.7\textwidth]{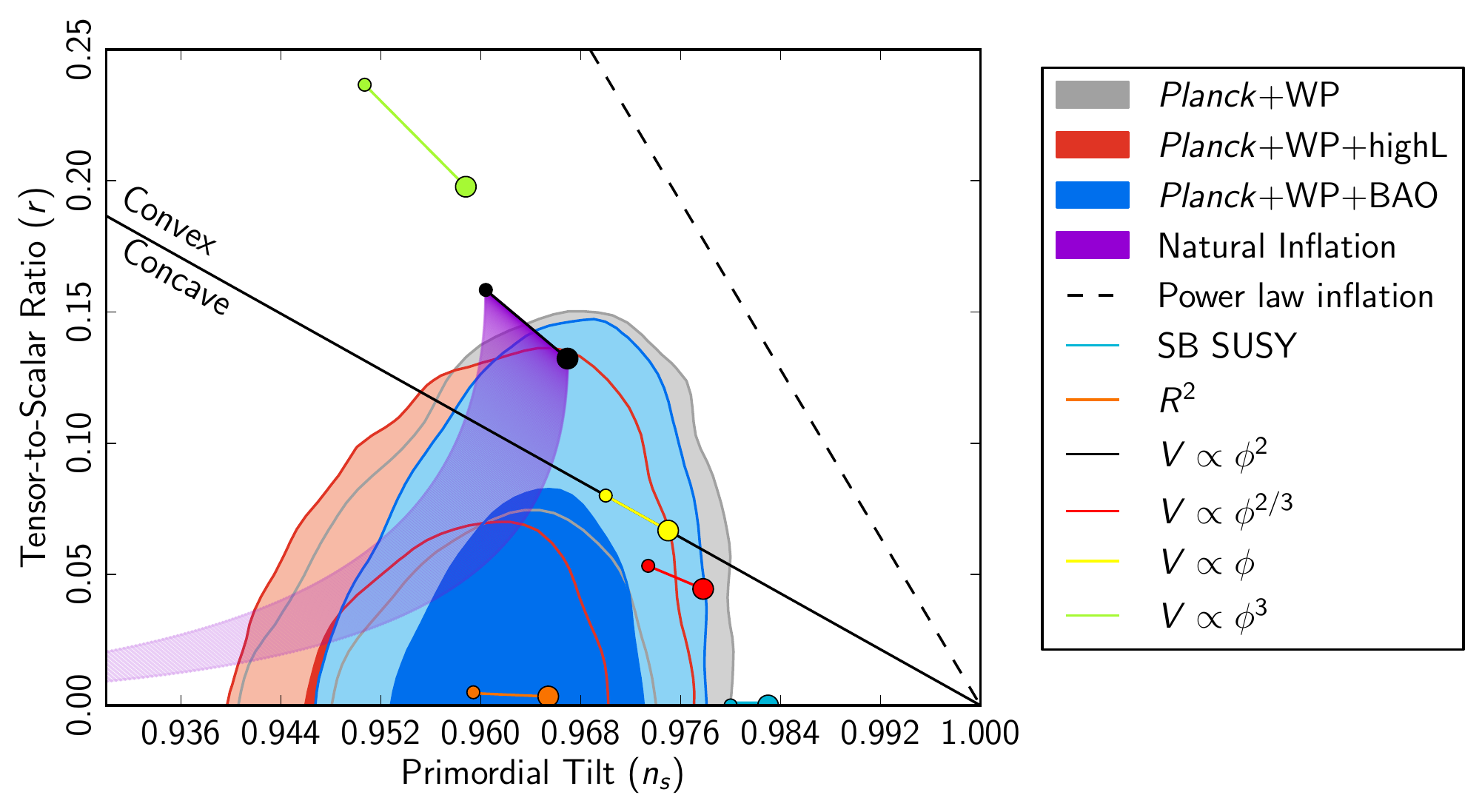}
    \caption{Planck 68\% and 95\% marginalised confidence levels for $n_s$ and $r$, taken from ref.~\protect\refcite{Planck}.}
    \label{fig:planck}
\end{figure}

Whilst models of this type (i.e. concave) are well supported by the Planck 2013 data~\cite{Planck}, they are amongst those disfavoured of the recent BICEP2 measurement of a large tensor to scalar ratio (\ref{large})~\cite{BICEP2}.
  This may be most simply seen from figure \ref{fig:planck}; for the Planck best fit value $n_s\sim0.96$, models with concave inflationary potentials are constrained to $r\lesssim0.11$. 

As mentioned above, there is a tension present between the BICEP2 and Planck results. Resolution of this issue is naturally outwith the scope of this article, so we instead focus on making contact between gravitino condensate inflation and precision inflationary phenomenology, so that once more data are available; hopefully reducing or eliminating the aforementioned tension, the viability of the model may be fully assessed.

It is important to stress however once more that this approach is only one of many methods of realising inflation in the context of supergravity. As already mentioned, both hybrid \cite{sugra_hybrid} and chaotic inflation \cite{sugra_chaotic} have been previously explored, and,
in light of the 2013 Planck results \cite{Planck}, a number of realisations of Starobinsky-type  inflation in supergravity have also been investigated \cite{sugra_staro,ahmstaro}.

The present article will review the dynamical breaking of SUGRA and its potential links to inflation, 
by discussing advantages and disadvantages of the various inflationary scenarios that are linked one way or another to gravitino condensation. Moreover, it 
will also deal with a number of important technical steps characterising the 
dynamical breaking of SUGRA which were not fully elaborated upon in ref.~\refcite{ahm}. 
Following some preliminaries establishing the framework within which we are working, we detail:
\begin{itemize}
	\item The nature of the Fierz ambiguity inherent to this approach, which affects the strength of the coupling into the scalar channel we are interested in, and our use of flat space Schwinger-Dyson equations to resolve the issue. 
	\item Computation of the $\langle\overline\psi_\mu\psi^\mu\rangle$ bound state propagator, yielding the wave function renormalisation $Z$ which controls the magnitudes of the various slow roll parameters, along with the overall energy scale of inflation.
\end{itemize}

The structure of the article is then as follows:
\begin{itemize}
	\item In Section \ref{sec:effpot} we review the formalism and physical concepts underlying dynamical breaking of SUGRA and the associated super-Higgs effect, within the context of simple 
four dimensional ${\mathcal N} =1$ models, including superconformal extensions thereof (with broken conformal symmetry) that are necessitated for phenomenological reasons, as explained in the text.  
	\item In Section \ref{sec:SD} we compute, within a (flat-space-time) Schwinger-Dyson formalism, the wave-function renormalisation of the gravitino condensate field. 
	\item We then extrapolate this wave-function renormalisation in Section \ref{sec:infl} to discuss hilltop inflation, where gravitino condensate fields near the origin of the effective potential play the role of the inflation field. The model is compatible with slow roll for very large values of the condensate wave function renormalisation.
	\item This prompts us to discuss in Section \ref{sec:star} alternative scenarios for inflation of Starobinsky type that may occur in the massive gravitino phase, near the non-trivial minimum of the effective potential. In such scenarios, which are compatible with the Planck but not the BICEP2 results, the role of the inflaton field is played by the scalar mode that describes the effects of scalar-curvature-square terms that characterise the gravitational sector of the effective action in the broken SUGRA phase, after integrating out the massive gravitinos. 
	\item Finally, conclusions are presented in section \ref{sec:concl}. Some technical aspects of our approach, associated with Fierz ambiguities in the SUGRA action, are discussed in an Appendix.
\end{itemize}

\section{Super-Higgs effect and dynamical breaking of ${\mathcal N}=1$ SUGRA at one loop \label{sec:effpot}} 

Our starting point is the $\mathcal{N}=1$ $D=4$ (on-shell) action for `minimal' Poincar\'e supergravity in the second order formalism, following the conventions of ref.~\refcite{Freedman} (with explicit factors of the (dimensionful) gravitational constant $\kappa^2 = 8\pi {\rm G} = 1/M_{\rm Pl}^2$, in units $\hbar=c=1$, where $M_{\rm Pl}$ the reduced Planck mass in four space-time dimensions):
 \begin{align}\label{sugraction}
	&S_{\rm{SG}}=\int d^4x \,e \left(\frac{1}{2\kappa^2}R\left(e\right)-\overline\psi_\mu\gamma^{\mu\nu\rho}D_\nu\psi_\rho+\mathcal{L}_{\rm torsion}\right),\\\nonumber
	&\kappa^2=8\pi G\,
	\quad\gamma^{\mu\nu\rho}=\frac{1}{2}\left\{\gamma^\mu,\gamma^{\nu\rho}\right\}\,,
	\quad \gamma^{\nu\rho}=\frac{1}{2}\left[\gamma^\nu,\gamma^\rho\right]\,,
\end{align}
where $R(e)$ and $D_\nu\psi_\rho\equiv\partial_\nu\psi_\rho+\frac{1}{4}\omega_{\nu ab}\left(e\right)\gamma^{ab}\psi_\rho$ are defined via the torsion-free connection and, given the gauge condition $\gamma\cdot\psi=0$, 
\begin{align}\label{torsion}
	\mathcal{L}_{\rm torsion}=-\frac{1}{16}\left(\left(\overline\psi^\rho\gamma^\mu\psi^\nu\right)\left(\overline\psi_\rho\gamma_\mu\psi_\nu+2\overline\psi_\rho\gamma_\nu\psi_\mu\right)\right)\times2\kappa^2\,,
\end{align}
arising from the fermionic torsion parts of the spin connection~\footnote{We note in passing that such four-fermion interactions are characteristic of any Einstein-Cartan theory of fermions in curved space-time~\cite{mercuri}. In fact, in a standard spin-1/2 fermion-gravity theory, the torsion-induced four fermion interactions assume a repulsive axial (pseudovector)-current-current form $-\left({\overline \psi }\gamma^\mu \gamma^5 \psi \right)\left({\overline \psi} \gamma_\mu \gamma^5 \psi \right)$. 
As we demonstrate in the Appendix and in section \ref{sec:fierz}, a corresponding repulsive axial-current-current term for the gravitino torsion terms can also be obtained by appropriately utilising Fierz identities in analogy with the Einstein-Cartan theory, cf. (\ref{couplings}).}.

Extending the action off-shell requires the addition of auxiliary fields to balance the graviton and gravitino degrees of freedom. 
These fields however are non-propagating and may only contribute to topic at hand through the development of scalar vacuum expectation values, which would ultimately be resummed into the cosmological constant.

Making further use of this gauge condition in concert with the Fierz identities (as detailed in the Appendix), we may write 
\begin{align}
		\mathcal{L}_{\rm torsion}
		=\lambda_{\rm S}\left(\overline\psi^\rho\psi_\rho\right)^2
		+\lambda_{\rm PS}\left(\overline\psi^\rho\gamma^5\psi_\rho\right)^2
		+\lambda_{\rm PV}\left(\overline\psi^\rho\gamma^5\gamma_\mu\psi_\rho\right)^2
\end{align}
where the couplings $\lambda_{\rm S}$, $\lambda_{\rm PS}$ and $\lambda_{\rm PV}$ express the freedom we have to rewrite each quadrilinear in terms of the others via Fierz transformation. 
This freedom in turn leads to a known ambiguity in the context of mean field theory \cite{Wetterich}, which we will address fully in section \ref{sec:fierz}.

Specifically, we wish to linearise these four-fermion interactions via suitable auxiliary fields, e.g.
\begin{align}
	\frac{1}{4}\left(\overline\psi^\rho\psi_\rho\right)^2\sim\sigma\left(\overline\psi^\rho\psi_\rho\right)-\sigma^2\,,
\end{align}	
where the equivalence (at the level of the action) follows as a consequence of the subsequent Euler-Lagrange equation for the auxiliary scalar $\sigma$.
Our task is then to look for a non-zero vacuum expectation value $\langle\sigma\rangle$ which would serve as an effective mass for the gravitino. This is however complicated by the fact that our coupling $\lambda_{\rm S}$ into this particular channel is, by virtue of Fierz transformations, ambiguous.

To induce the super-Higgs effect~\cite{DeserZumino} we also couple in the Goldstino associated to global supersymmetry breaking via the addition of
\begin{align}\label{goldstino}
	\mathcal{L}_\lambda=f^2\det\left(\delta_{\mu\nu}+\frac{i}{2f^2}\overline\lambda\gamma_\mu\partial_\nu\lambda\right)\bigg|_{\gamma\cdot\psi=0}=f^2+\dots\,,
\end{align}	
where $\lambda$ is the Goldstino, $\sqrt{f}$ expresses the scale of global supersymmetry breaking, and \dots\, represents higher order terms which may be neglected in our weak-field expansion of the determinant.
It is worth emphasising at this point the universality of \eqref{goldstino}; any model containing a Goldstino may be related to $\mathcal{L}_\lambda$ via a non-linear transformation \cite{Komargodski}, and thus the generality of our approach is preserved. 

Upon a specific gauge choice for the gravitino field
$$ \gamma^\mu \psi_\mu = 0~,$$ 
and an appropriate redefinition, one may eliminate any presence of the Goldstino field from the final effective 
action describing the dynamical breaking of local supersymmetry, except the cosmological constant term $f^2$ in (\ref{goldstino}), which serves as a reminder of the pertinent scale of supersymmetry breaking.
The non-trivial energy scale this introduces, along with the disappearance (through field redefinitions) of the Goldstino field from the physical spectrum and the concomitant development of a gravitino mass, characterises the super-Higgs effect. 

We may then identify in the broken phase an effective action
\begin{align}\label{finalaction}
	S=\frac{1}{2\kappa^2}\int d^4x \,e \left(\left(R\left(e\right)-2\Lambda\right)-\overline\psi_\mu\gamma^{\mu\nu\rho}D_\nu\psi_\rho+ m_{\rm dyn} \, \left(\overline\psi_\mu \psi^\mu\right)\right),	
\end{align}
where $\Lambda$ is renormalised cosmological constant, to be contrasted with the (negative) tree level cosmological constant 
\begin{align}
	 \Lambda_0\equiv\kappa^2\left(\sigma ^2  -f^2\right)~,
\end{align}
and $m_{\rm dyn} \propto \langle \sigma \rangle $ is a dynamically generated gravitino mass, the origin of which will be explained presently.
It is worth stressing at this point that $\Lambda_0$ must be negative due to the incompatibility of supergravity with dS vacua; if SUGRA is broken at tree level, then of course no further dynamical breaking may take place.

For phenomenological reasons which will be outlined below, as in refs.~\refcite{emdyno,ahm} we adopt an extension of ${\mathcal N}=1$ SUGRA which incorporates local supersymmetry in the Jordan frame, enabled by an associated dilaton superfield~\cite{confsugra}. 
The scalar component $\varphi$ of the latter can be either a fundamental space-time scalar mode of the gravitational multiplet, i.e. the trace of the graviton (as happens, for instance, in supergravity models that appear in the low-energy limit of string theories), or a composite scalar field constructed out of matter multiplets.
In the latter case these could include the standard model fields and their superpartners that characterise the Next-to-Minimal Supersymmetric Standard Model, which can be consistently incorporated in such Jordan frame extensions of SUGRA~\cite{Ferrara:2010in}. 

Upon appropriate breaking of conformal symmetry, induced by specific dilaton potentials (which we do not discuss here), one may assume that the dilaton field acquires a non-trivial vacuum expectation value $\langle \varphi \rangle \ne 0 $. 
One consequence of this is then that in the broken conformal symmetry phase, the resulting supergravity sector, upon passing (via appropriate field redefinitions) to the Einstein frame is described by an action of the form (\ref{sugraction}), but with the coupling of the gravitino four-fermion interaction terms being replaced by 
\begin{equation}\label{tildcoupl}
\tilde \kappa\equiv e^{-\langle\varphi\rangle}\kappa~,
\end{equation}
while the Einstein term in the action carries the standard gravitational coupling $1/2\kappa^2$. 

\begin{figure}[h!!]
		\centering
		\includegraphics[width=0.7\textwidth]{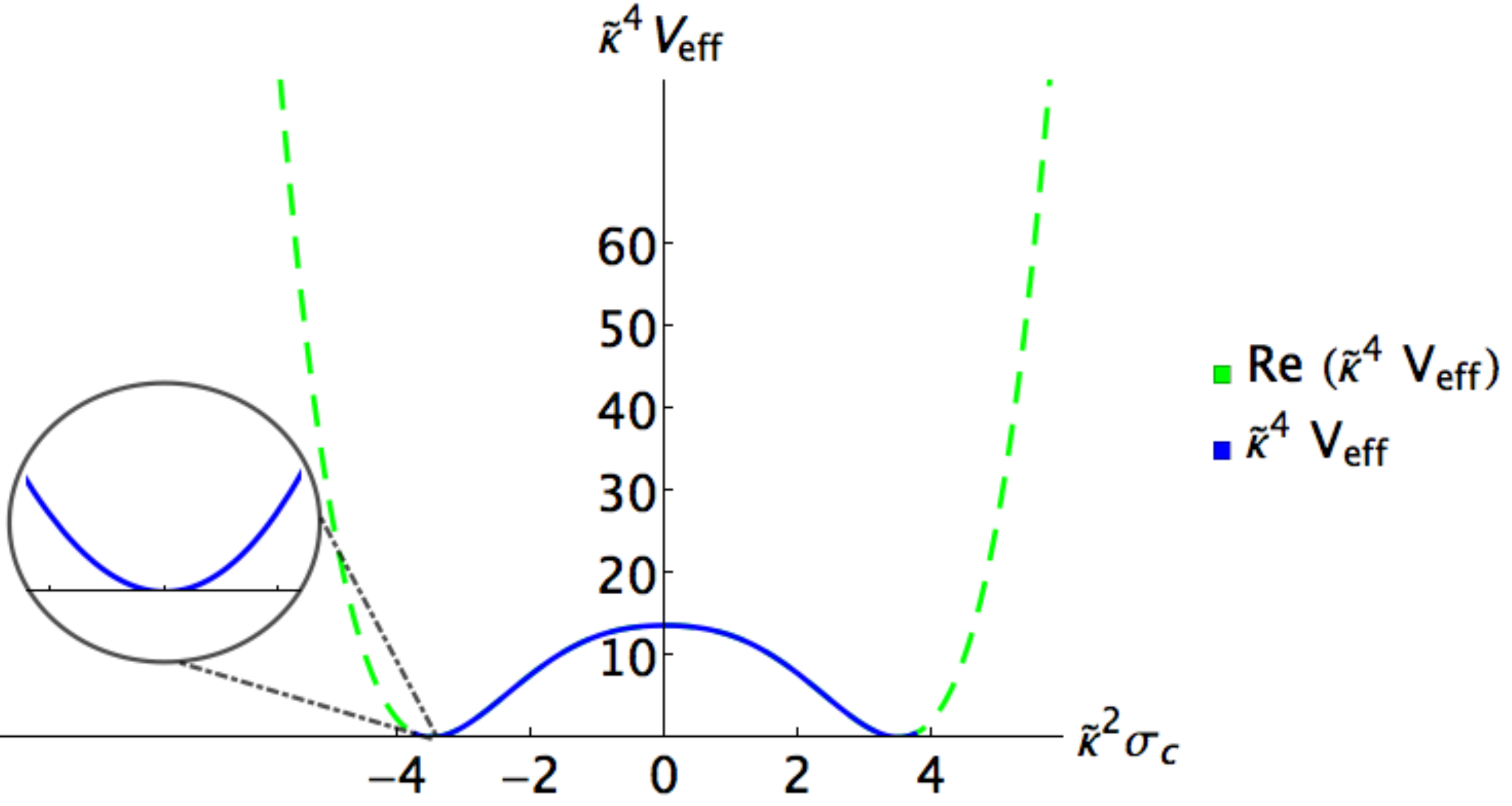} \vfill
		\includegraphics[width=0.4\textwidth]{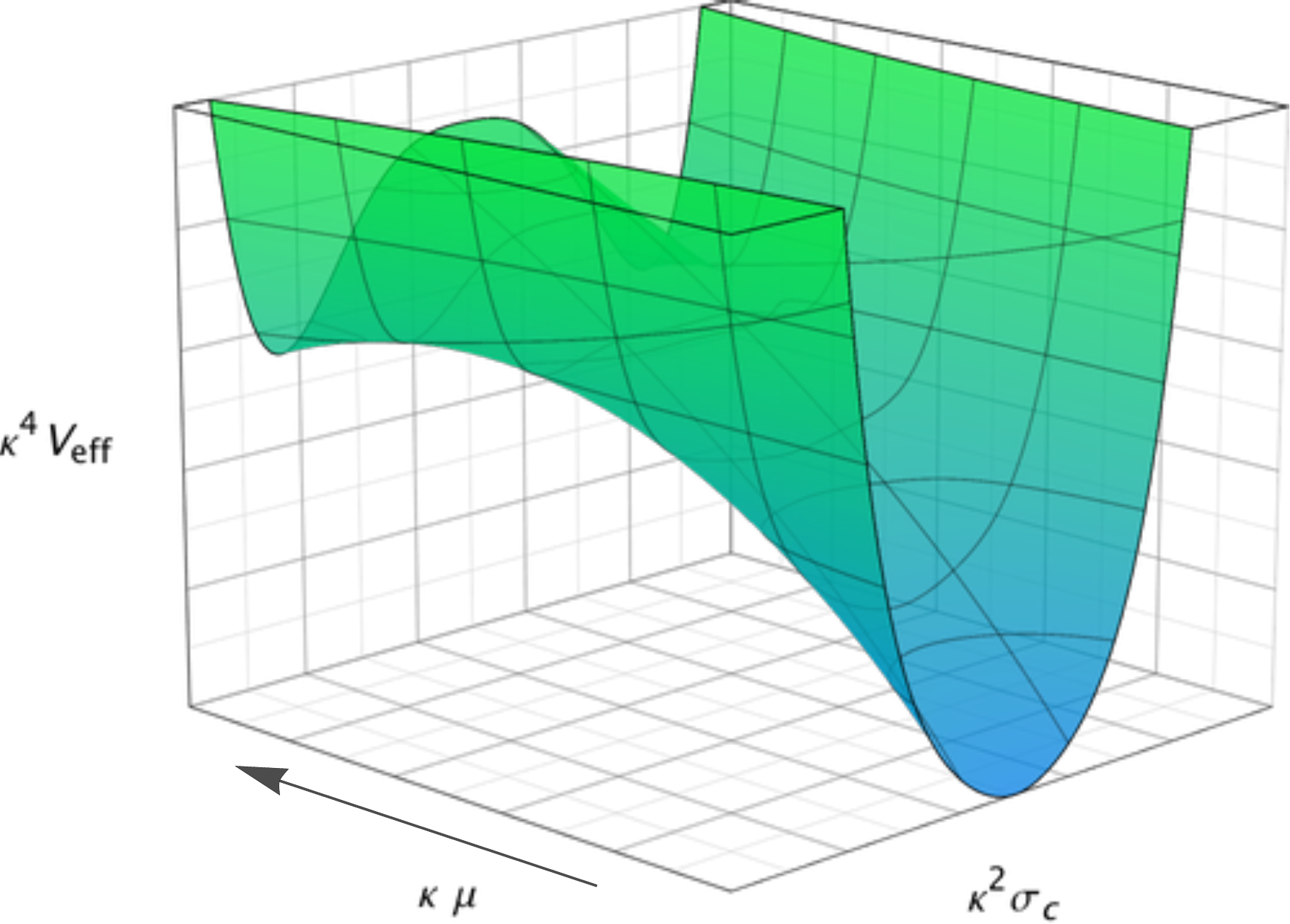}
		\includegraphics[width=0.4\textwidth]{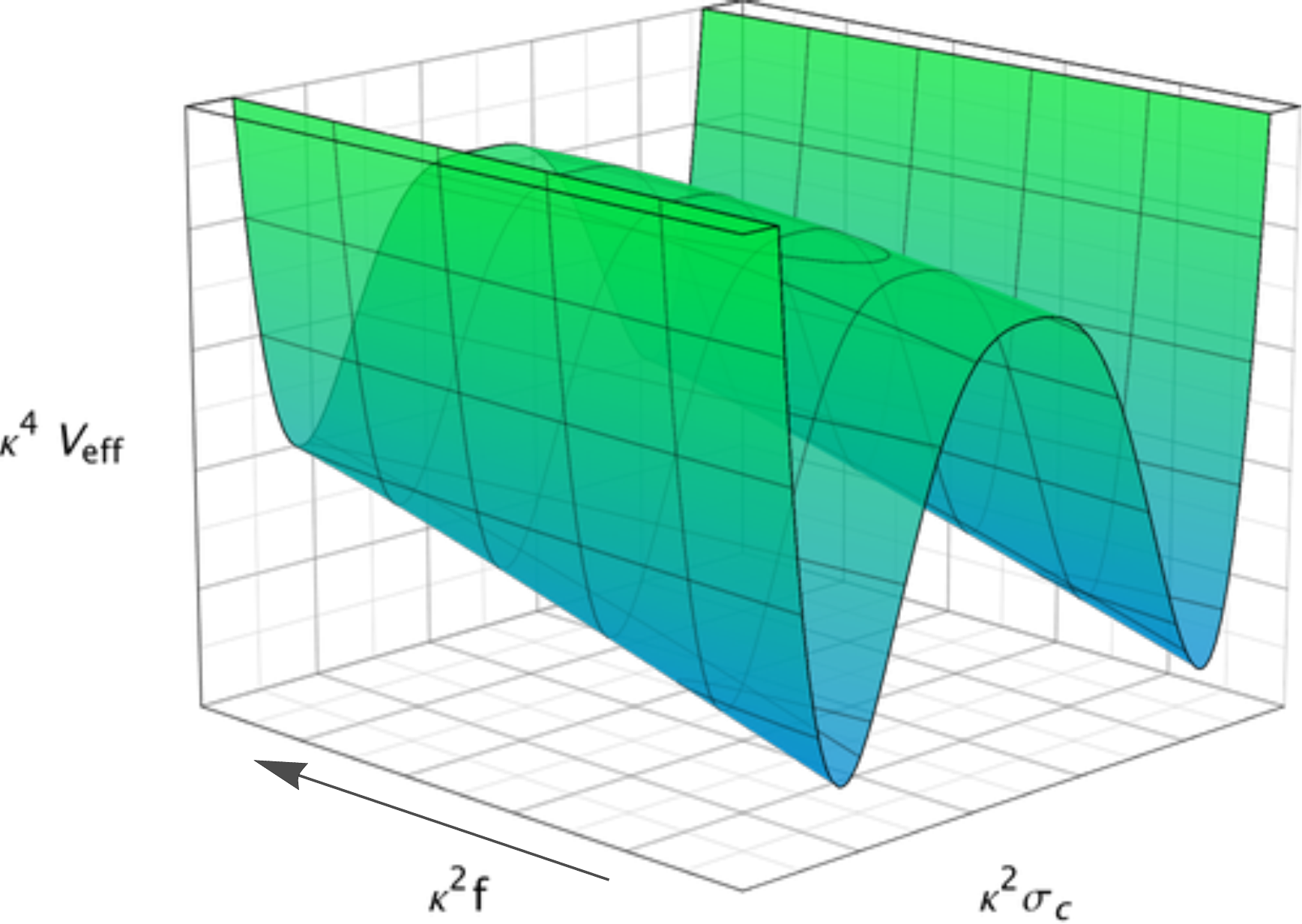}
		\caption{\emph{Upper panel}: The effective potential \eqref{effpotsugra}, expressed in units of the 
		coupling $\tilde \kappa$ (\ref{tildcoupl}). \emph{Lower panel}: As above, but showing schematically the effect of tuning the  RG scale $\mu$ and the supersymmetry breaking scale $f$, whilst holding, respectively, $f$ and $\mu$ fixed. The arrows in the respective axes correspond to the direction of increasing $\mu$ and $f$.}
\label{fig:effpotsugra}	
	\end{figure}

Expanding the graviton field about a de Sitter background~\cite{fradkin} (under the assumption that it is a solution of the one-loop effective equations) with renormalised cosmological constant $\Lambda > 0$, and integrating out both bosonic and fermionic quantum fluctuations to one loop yields the following effective potential for the gravitino condensate field $\sigma$ in the flat space-time limit $\Lambda \to  0$, as detailed in ref.~\refcite{ahm}, 
\begin{align}\label{effpotsugra}
		V_{\text{eff}}=V_{B}^{(0)}+V_{B}^{(1)}+V_{F}^{(1)}
		=-\frac{\Lambda_0}{\kappa^2}+V_{B}^{(1)}+V_{F}^{(1)}~,
		\quad \Lambda_0\equiv\kappa^2\left(\sigma^2-f^2\right)~,
	\end{align}
where 	\begin{align}\label{boson}
		V_{B}^{(1)}=\frac{45 \kappa ^4}{512 \pi^2}\left(f^2-\sigma ^2\right)^2 \left(3-2 \ln \left(\frac{3 \kappa ^2 \left(f^2-\sigma ^2\right)}{2 \mu ^2}\right)\right)~, 
		\end{align} 
and 		\begin{align}\label{fermion}
		&V_{F}^{(1)} =\frac{{\tilde \kappa} ^4 \sigma^4}{30976 \pi ^2} \left(30578 \ln \left(\frac{{\tilde \kappa}^2 \sigma^2}{3 \mu^2}\right)-45867+29282 \ln \left(\frac{33}{2}\right)+1296 \ln\left(\frac{54}{11}\right)\right) \nonumber \\ 
& = \left(\frac{{\tilde \kappa}}{\kappa}\right)^4 \, \frac{\kappa^4 \sigma^4}{30976 \pi ^2} \left(30578 \ln \left(\left(\frac{{\tilde \kappa}}{\kappa}\right)^2\, \frac{\kappa^2 \sigma^2}{3 \mu ^2}\right)-45867+29282 \ln \left(\frac{33}{2}\right)+1296 \ln\left(\frac{54}{11}\right)\right)~,
	\end{align}
indicate the contributions to the effective potential from bosonic and fermionic fields respectively, and 	
$\mu$ is an inverse renormalisation group (RG) scale.
The effective potential (\ref{effpotsugra}) is depicted in fig.~\ref{fig:effpotsugra}. 

We may firstly note that as we flow from UV to IR (\emph{i.e}. in the direction of increasing $\mu$),  we obtain the correct double-well shape required for the super-Higgs effect, and secondly that tuning $f$ allows us to shift $V_{\text{eff}}$ and thus attain the correct vacuum structure (\emph{i.e.} non-trivial minima $\sigma_c$ such that $V_{\text{eff}}\left(\sigma_c\right)=0$).
Moreover,  the shape of the effective potential changes, as one varies the 
(renormalisation) scale $\mu$ from ultraviolet to infrared values (\emph{i.e.} flowing in the direction of increasing $\mu$), in such a way that the broken symmetry phase (double-well shaped potential) is reached in the IR. 
This indicates that the dynamical generation of a gravitino mass is actually an IR phenomenon, in accordance with the rather general features of dynamical mass in field theory. 

In the broken phase, the mass of the gravitino condensate is then given by 
\be
m_\sigma^2\equiv V''_{\text{eff}}(\sigma_c)~,
\ee
where $\sigma_c$ is the minimum of $V_{\text{eff}}$ and a prime denotes a functional derivative with respect to the gravitino-condensate field. As observed from (\ref{boson}), the bosonic contributions to the effective potential contain logarithmic terms which would contribute imaginary terms, leading to instabilities, unless 
\be\label{imagin}
\sigma_c^2<f^2~.
\ee
From \eqref{effpotsugra} it is straightforward to see that this condition is equivalent to the negativity of the tree-level cosmological constant $\Lambda_0$, which is entirely sensible; if $\Lambda_0>0$ then SUGRA is broken at tree level (given the incompatibility of supersymmetry with de Sitter vacua) and there can be no dynamical breaking.
As such, we must then tune $f$ for a given value of $\mu$ to find self consistent minima $\sigma_c$ satisfying (\ref{imagin}), thereby ensuring a real $V_{\text{eff}}$. In fact, here lies the importance of the super-Higgs effect, and thus of a non-zero positive $f^2 > \sigma_c^2 > 0$, in allowing dynamical breaking of local supersymmetry~\footnote{It should be mentioned at this point that in refs.~\refcite{odintsov}, the importance of the super-Higgs effect was ignored, which led to the incorrect conclusion that imaginary parts exist necessarily in the one-loop effective potential (in the same class of gauges as the one considered in ref.~\refcite{ahm} and here) and hence dynamical breaking of SUGRA was not possible. As we have seen above, such imaginary parts are absent when the condition (\ref{imagin}) is satisfied, and thus dynamical breaking of SUGRA occurs.}. 

As discussed in refs.~\refcite{emdyno,ahm}, phenomenologically realistic situations, where one avoids transplanckian gravitino masses, for supersymmetry breaking scales $\sqrt{f} $ at most of order of the Grand Unification (GUT) scale $10^{15-16}$~GeV, as expected from arguments related to the stability of the electroweak vacuum, can occur only for 
large $\tilde \kappa$ couplings, typically of order $\tilde \kappa \sim \Big(10^{3}-10^{4} \Big)\, \kappa$. 
Given the relation \eqref{tildcoupl} this corresponds to dilaton vev of $\mathcal{O}\left(-10\right)$, where the negative sign may be familiar in the context of dilaton-influenced cosmological scenarios \cite{Antoniadis:1988aa}.

If we consider for concreteness the case $\tilde\kappa=10^3 \kappa$, which is a value dictated by the inflationary phenomenology of the model~\cite{emdyno}, we may find solutions with a vanishing one-loop effective potential at the non-trivial minima corresponding to:
\begin{eqnarray}\label{solutions}
{\tilde \kappa}^2 \, \sigma_c \simeq 3.5~, \quad {\tilde \kappa}^2 \, f \simeq 3.7~, \quad {\tilde \kappa}\, \mu \simeq 4.0~, 
\end{eqnarray}
which leads to a global supersymmetry breaking scale 
	\begin{equation}\label{fscaleconf}
		\sqrt{f} \simeq 4.7\times10^{15}~{\rm GeV}~,
	\end{equation}
 and dynamical gravitino mass 
 	\begin{equation}\label{gravinoconf}
 		m_{\rm dyn}  
		\simeq 2.0\times10^{16}~{\rm GeV}~.
	\end{equation}
	
At the non-trivial minima we find $\tilde\kappa^4V_{F}^{(1)}\simeq-1.4$, $\tilde\kappa^4V_{B}^{(1)}\simeq5.9\times10^{-13}$, with tree-level cosmological constant $\tilde\kappa^2\Lambda_0 \simeq-1.4$.
We thus observe that fermion contributions to the effective potential are much stronger than the corresponding bosonic contributions for the cases of large couplings $\tilde \kappa \gg \kappa$. 
These values are phenomenologically realistic, thereby pointing towards the viability (from the point of view of producing realistic results of relevance to phenomenology) of the scenarios of dynamical breaking of local supersymmetry in conformal supergravity models. 

On the other hand, in standard SUGRA scenarios, where $\tilde \kappa = \kappa$, one finds, as already mentioned, transplanckian values for the dynamically generated gravitino mass~\cite{ahm}:
$m_{\rm dyn} \simeq 2.0\times10^{19}~{\rm GeV}$, and a 
global supersymmetry breaking scale $\sqrt{f} \simeq 4.7\times10^{18}~{\rm GeV}$, far too high
to make phenomenological sense.

In order to discuss the possible connection with inflation, we need to calculate one more important ingredient; the wave-function renormalisation. In principle, this should be calculated in a curved de Sitter space-time, which characterises the (unbroken) phase of SUGRA, when the condensate field is near the trivial maximum of the effective potential (\ref{effpotsugra}). 
This is a complicated task and will shall not be presented here. 
However, it turns out that, since, according to the data~\cite{Planck,encyclo}, the de Sitter phase Hubble parameter in phenomenologically relevant inflationary models is expected to be several orders of magnitude smaller than the Planck scale, $m_P$ (\ref{HI}),
the space-time curvature during inflation is not too large, and thus a flat space-time estimate of the wave function renormalisation may suffice.
In the scenario of \cite{ahm}, such an estimate characterises the broken SUGRA phase at the end of inflation. 
This will be the topic of the next section. 

A further extension of this flat-space analysis has also been performed in \cite{ahmstaro}, with the conclusion that it is possible to have a second inflationary phase of Starobinsky type~\cite{staro}, succeeding the hilltop inflation, if the latter exists. 
We shall discuss this case in section \ref{sec:star}. 
This Starobinksy-type inflation appears to be more natural than the hilltop inflation, in the context of the dynamically broken SUGRA, in the sense that it is not characterised by unnaturally large parameters. 
Nevertheless, it leads to much more suppressed values of the tensor-to-scalar ratio for the primordial fluctuations, which although in agreement with Planck results~\cite{Planck}, are in stark disagreement with BICEP2~\cite{BICEP2}. 

\section{Schwinger-Dyson Gravitino Mass Generation in flat space-time \label{sec:SD}}

\subsection{Gap Equation} 

As we discussed above, the gravitino torsion parts of the effective SUGRA lagrangian (\ref{sugraction}),(\ref{torsion}), 
contain four-fermion interactions, and thus one is facing a situation similar to that of the dynamical chiral symmetry breaking 
of the Nambu-Jona-Lasinio model~\cite{NJL}.  
Following the analysis of ref.~\refcite{ahm}, in this section we shall discuss the generation of a gravitino mass within the context of a Scwinger-Dyson (SD) approach in flat space-time backgrounds, which, as already mentioned, may be viewed as the final stage of the inflationary scenario of ref.~\refcite{ahm}.  
This formalism allows for an estimate of the wave-function renormalisation of the gravitino condensate quantum field, which is essential for the inflationary phenomenology to be discussed in subsequent sections. 

For our SD analysis below we shall need the propagator for the massive gravitino in flat space time, which reads~\cite{Nieuwenhuizen}
\be
P_{\mu\nu}=-\frac{i}{2}\gamma_\mu\frac{\br p+m_{\rm dyn}}{p^2-m_{\rm dyn}^2}\gamma_\nu~,
\ee
where $m_{\rm dyn}$ is the gravitino mass.

For our current purposes we note that, in a Hartree-Fock approximation, according to which one identifies the gravitino mass with the scalar condensate,
$m_{\rm dyn}$ is a solution of the gap equation 
\bea\label{gap}
m_{\rm dyn}&=&-\frac{\lambda_{\rm S}}{2}\lim_{x\to0}\Tr\left(P_{\mu\nu}(x)\right)
=8\lambda_{\rm S} i\int \frac{d^4p}{(2\pi)^4}\frac{m_{\rm dyn}}{p^2-m_{\rm dyn}^2}~,
\eea
where the dimensionful coupling $\lambda_{\rm S}$ is not fixed.
The right-hand side of \eqref{gap} is represented by a quadratically divergent tadpole diagram, yielding
\be
m_{\rm dyn}=\frac{\lambda_{\rm S} m_{\rm dyn}}{2\pi^2}\left(C_{\rm off}^2-m_{\rm dyn}^2\ln\left(\frac{C_{\rm off}^2}{m_{\rm dyn}^2}\right)\right)~,
\ee
regulated by a (flat space) cut off $C_{\rm off}$.

Since $m_{\rm dyn}<C_{\rm off}$, we can see that, if the dimensionful coupling is too small
\be
\lambda_{\rm S}<\lambda_{\rm S}\big|_{\rm crit.}=\frac{2\pi^2}{C_{\rm off}^2}~,
\ee
then the only solution to the gap equation \eqref{gap} is $m_{\rm dyn}=0$. On the other hand, if $\lambda_{\rm S}>\lambda_c$, 
then the gap equation (\ref{gap}) has a non-trivial solution $\omega\equiv m_{\rm dyn}/C_{\rm off}$, which satisfies 
\be\label{gap2}
\omega^2\ln(\omega^2)=\frac{1}{g}-1~<0~~,~~~\mbox{with}~~g\equiv\frac{\lambda_{\rm S} C_{\rm off}^2}{2\pi^2}>1~.
\ee
But one can also see that, if $1/g-1<-e^{-1}$, the the gap equation (\ref{gap2}) has no solution. 
Therefore, a non-trivial dynamical mass 
implies that the dimensionless coupling constant $g$ satisfies
\be\label{finetuning}
1<g\leq\frac{1}{1-e^{-1}}\simeq1.58~,
\ee
which will be assumed in the following.

We may solve \eqref{gap2} exactly via the Lambert W-function, which is defined as the set of functions $W$ for which 
\begin{align}	
	z=W\left(z\right)e^{W\left(z\right)}\, \forall z \in \mathbb{C}\, ,
\end{align}
yielding the relation
	\begin{align}\label{musolution}
		\omega^2=e^{W\left(g^{-1}-1\right)}\,.
	\end{align}
where we note that since \eqref{gap2} admits multiple solutions for a given value of $g$ (e.g. for $g\to1$, we may have $\omega\to1$ or $\omega\to0$), \eqref{musolution} must also necessarily be multivalued.
We may formalise this by considering both the principal and lower branches of $W\left(z\right)$, denoted $W_{0}\left(z\right)$ and $W_{-1}\left(z\right)$ respectively. 
As we will see in the following however, only the lower branch is of phenomenological interest.
\begin{figure}[h!]
  \centering
    \includegraphics[width=0.5\textwidth]{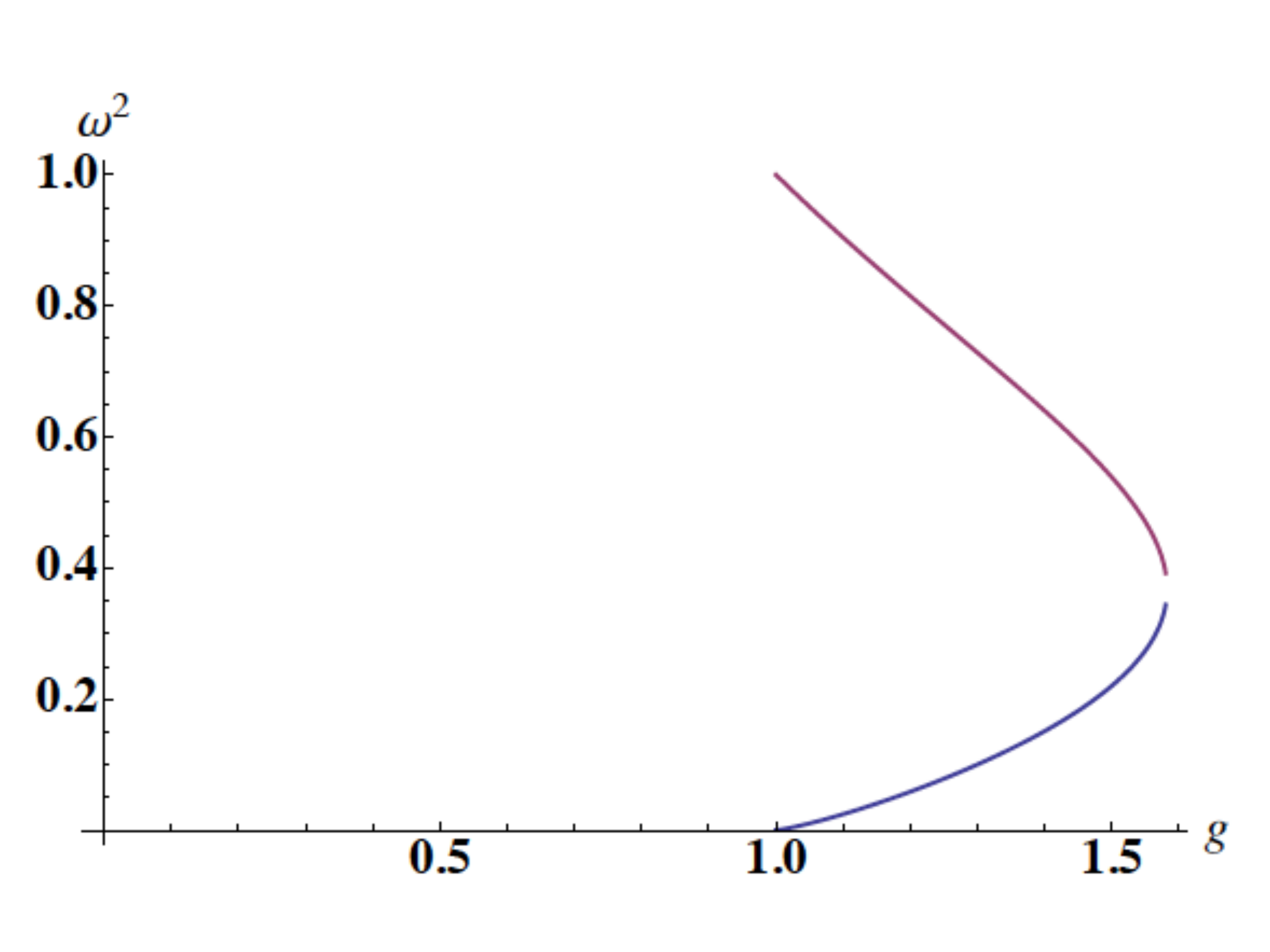}
    \caption{Both branches of \eqref{musolution}, the upper curve being the principal branch}
    \label{fig:muplot}
\end{figure}

We also note that, for the dynamical mass to be small compared to the cut off, it is necessary that $g\simeq1$, or equivalently that $\lambda_{\rm S}\simeq \lambda_{\rm S}\big|_{\rm crit.}$;
this fine tuning is equivalent to the hierarchy problem in the Standard Model \cite{Bardeen}. 

\subsection{Fierz ambiguity \label{sec:fierz}}

As previously stated, we wish to linearise the four-fermion interactions via suitable auxiliary fields, e.g.	
\begin{align}
	\frac{1}{4}\left(\overline\psi^\rho\psi_\rho\right)^2\sim\sigma\left(\overline\psi^\rho\psi_\rho\right)-\sigma^2\,,
\end{align}	
where a non-zero scalar vacuum expectation value $\langle\sigma\rangle$ would then source the $m_{\rm dyn}$ of the previous section. 
By virtue of Fierz transformations however, the coupling $\lambda_{\rm S}$ into this particular channel is ambiguous; we may always transform the left hand side into a pseudovector or pseudoscalar before linearising, conceivably yielding no scalar condensate at all, or possibly additional (and unwanted) pseudovector and pseudoscalar condensates.

We may understand the ambiguity as arising from this linearisation (or equivalently mean-field theory in general), which distributes of the original four-fermion interaction into (presumed to be independent) scalar, pseudoscalar and pseudovector channels~\cite{Wetterich}. To know the actual relative magnitudes of coupling into these channels concretely would require knowledge beyond the pointlike limit, and thus beyond the (perturbative) supergravity approximation.

An exact renormalisation-group analysis, along the lines of standard NJL models~\cite{Wetterich}, may allow this issue to be addressed, and, in the specific case of SUGRA, an embedding within some string theory model should also offer some resolution.
We will however proceed within the framework of perturbation theory.

To address this issue perturbatively in our formalism, we may note firstly that the three couplings $\lambda_{\rm S}$, $\lambda_{\rm PS}$ and $\lambda_{\rm PV}$ only span a two dimensional parameter space,
since they must resum to yield the (unambiguous) expression \eqref{torsion}.
This gives the relation
\begin{align}\label{couplings}
	\left(\lambda_{\rm S}-\lambda_{\rm PS}+4\lambda_{\rm PV}\right)
	=-\frac{3}{8}\times2\kappa^2\,,
\end{align}
from which we may see that e.g. if we rewrite all pseudoscalars and pseudovectors as scalars (thus giving zero coupling into those channels) then $\lambda_{\rm S}=-3/8 \times2\kappa^2$.

To fully circumvent the ambiguity we may derive two further constraints on these couplings in flat space, which we may then assume to hold in generality. 

As a first condition, it seems plausible to require that looking only in the scalar channel we should find a suitable nonzero vacuum expectation value $\langle\sigma\rangle$.
Given that we are interested in a non-zero and phenomenologically desirable gravitino mass (i.e. $0<m_{\rm dyn}/M_{\rm P}<<1$), this provides a first constraint from the results of the previous section
\begin{align}\label{scalar coupling}
	\lambda_{\rm S}\simeq	\lambda_{\rm S}\big|_{\rm crit.}=\frac{2\pi^2}{C_{\rm off}^2}\,.
\end{align}

As a second condition, we may consider the lowest order Schwinger-Dyson equation
\begin{align}\label{SD}
	G_{\rm F}^{-1}=G_{\rm F0}^{-1}+\Sigma_{\rm F}
\end{align}
where $G_{\rm F}$ is the full fermion propagator, $G_{\rm F0}$ the free propagator, and $\Sigma_{\rm F}$ the self-energy.
\begin{figure}[h!]
  \centering
      \includegraphics[width=0.5\textwidth]{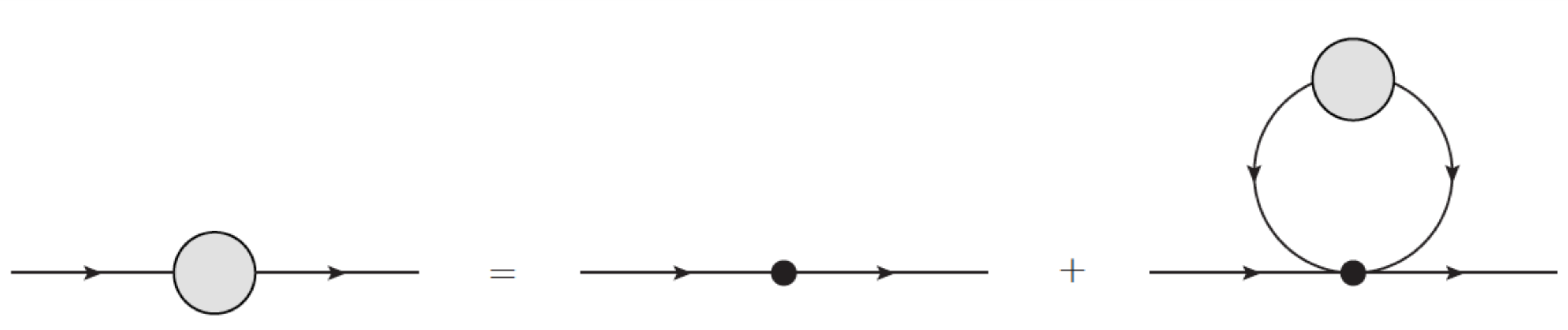}
    \caption{Schwinger-Dyson equation corresponding to \eqref{SD}.}
    \label{fig:SD graphs}
\end{figure}

This implies a gap equation for $\Sigma_{\rm F}$ as a function of the mass $m_{\rm dyn}$, which in turn must satisfy an analogous gap equation to \eqref{gap}, with the form
\begin{align}\label{SD1}
	m_{\rm dyn}=\frac{\left(\lambda_{\rm S}+\lambda_{\rm PS}-\lambda_{\rm PV}\right)}{2\pi^2}\int^{C_{\rm off}}_0 p^3 dp \frac{m_{\rm dyn}}{p^2+m_{\rm dyn}^2} \,,
\end{align}
where we are necessarily summing over all three channels as they all contribute to the self energy, and the relative sign difference of $\lambda_{\rm PV}$ arises from simple anti-commutation of gamma matrices.
The requirement again of nontrivial and phenomenologically desirable (i.e. $0<m_{\rm dyn}/M_{\rm P}<<1$) solutions to this equation then provides an analogous relation for the couplings into all three channels
\begin{align}\label{critical coupling}
	\left(\lambda_{\rm S}+\lambda_{\rm PS}-\lambda_{\rm PV}\right)\simeq
	\left(\lambda_{\rm S}+\lambda_{\rm PS}-\lambda_{\rm PV}\right)\big|_{\rm crit.}
	=\frac{2\pi^2}{C_{\rm off}^2}\, ,
\end{align}
which we may assume to be satisfied.
With three relations to be satisfied in three variables, this then removes the last vestiges of Fierz ambiguity by fixing the couplings to suitable values.

We may conclude from \eqref{scalar coupling} and \eqref{critical coupling} that $\lambda_{\rm PS}\simeq\lambda_{\rm PV}$ and then from \eqref{couplings} that $\lambda_{\rm PS}\simeq-(\kappa^2/4)-(2\pi^2/3C_{\rm off}^2)$. 
Given these reasonable criteria we then have the favourable scenario of an attractive coupling into the scalar channel, and repulsive couplings into the pseudoscalar and pseudovector channels, eliminating the possibility of undesirable pseudoscalar or pseudovector condensates.

Furthermore, identifying $C_{\rm off}$ with the Planck scale (as is natural in flat space) the numerical value of the scalar coupling given by \eqref{scalar coupling} is close to the value of $11\kappa^2/16$ used in ref.~\refcite{ahm}, where the Fierz ambiguity had not yet been addressed. 
This confirms the validity of the results therein in light of the issue raised by the Fierz ambiguity, and permits their straightforward reuse.

\subsection{Wave function renormalisation} 
 
The aim of this section is to derive the wave function renormalisation $Z$ of the gravitino bound state $\langle\psi_\mu\psi^\mu\rangle$, 
for which we will detail the steps followed by the authors of ref.~\refcite{Bardeen}, based on the usual Nambu and Jona-Lasinio approach \cite{NJL} that 
we review here.

The dynamical mass $m_{\rm dyn}$ found in ref.~\refcite{ahm} is proportional to the minimum of the effective potential $V(\sigma)$ 
for the auxiliary field $\sigma$, obtained after integration of the graviton and gravitino degrees of freedom.
In the present context, the Nambu and Jona-Lasinio gap equation will play the role of an effective approach, where the couping constant and the cut off 
are free parameters. Assuming that $m_{\rm dyn}$ satisfies 
this gap equation, we will be able to derive a consistent expression for $Z$, which will depend on an effective dimensionless coupling only.

The existence of the bound state can be described by the Bethe-Salpeter equation, involving the scalar bound state propagator $\Gamma$ (see ref.~\refcite{Hoyer} for a recent and pedagogical review in the context of gauge theories). 
This self-consistent equation can be expressed in the ladder/rainbow approximation as
\be
\Gamma=-\frac{\lambda}{2}+\frac{\lambda}{2}\Tr\int P_{\mu\nu}\Gamma P^{\nu\mu}~,
\ee
and leads, by iteration, to a geometric series of bubble graphs $B=-(\lambda/2)\Tr\int P_{\mu\nu}P^{\nu\mu}$, 
which can be resummed as 
\be\label{BS}
\Gamma=-\frac{\lambda}{2}\left(1+B+B^2+B^3+\cdots\right)=-\frac{\lambda/2}{1-B}~.
\ee
Each bubble graph is calculated for both particles with momentum $k/2$, where $k$ is the centre of mass momentum, which leads to
\bea
B(k)&=&i\frac{\lambda}{2}~\Tr\int\frac{d^4p}{\left(2\pi\right)^4}\frac{\gamma_\mu\left(\br p-\br k/2-m_{\rm dyn}\right)\gamma_\nu\gamma^\nu\left(\br p+\br k/2-m_{\rm dyn}\right)\gamma^\mu}{\left[\left(p-k/2\right)^2-m_{\rm dyn}^2\right]\left[\left(p+k/2\right)^2-m_{\rm dyn}^2\right]}\\
&=&2i\lambda\int\frac{d^4p}{\left(2\pi\right)^4}\frac{4p^2-k^2+4m_{\rm dyn}^2}{\left[\left(p-k/2)^2-m_{\rm dyn}^2\right]\left[\left(p+k/2\right)^2-m_{\rm dyn}^2\right)\right]}\nn
&=&2i\lambda\int\frac{d^4p}{\left(2\pi\right)^4}\frac{4\left[\left(p-k/2\right)^2-m^2\right]+2\left(4m_{\rm dyn}^2-k^2\right)+4k_\mu p^\mu}
{\left[\left(p-k/2\right)^2-m_{\rm dyn}^2\right]\left[\left(p+k/2\right)^2-m_{\rm dyn}^2\right]}\nn
&=&8i\lambda\int\frac{d^4p}{\left(2\pi\right)^4}\frac{1}{p^2-m_{\rm dyn}^2}
+4i\lambda\int\frac{d^4p}{\left(2\pi\right)^4}\frac{\left(4m_{\rm dyn}^2-k^2\right)}{\left(p^2-m^2\right)\left[\left(p+k\right)^2-m_{\rm dyn}^2\right]}~.\nonumber
\eea
As a result of the resummation \eqref{BS}, we obtain then
\be
\Gamma\left(k\right)=-\frac{\lambda}{2}\left[1-8i\lambda\int \frac{d^4p}{\left(2\pi\right)^4}\frac{1}{p^2-m_{\rm dyn}^2}
-4i\lambda\int \frac{d^4p}{\left(2\pi\right)^4}\frac{\left(4m_{\rm dyn}^2-k^2\right)}{\left(p^2-m_{\rm dyn}^2\right)\left[\left(p+k\right)^2-m_{\rm dyn}^2\right]}\right]^{-1}~,
\ee
where the first two terms cancel each other, if one assumes the gap equation (\ref{gap}) to be satisfied for $m_{\rm dyn}\ne0$. 
As a consequence, no quadratic divergence appears explicitly in the propagator, which can be expressed as
\bea\label{Gammap}
\Gamma(k)&=&\frac{\lambda}{2}\left[4i\lambda
\int \frac{d^4p}{\left(2\pi\right)^4}\frac{\left(4m_{\rm dyn}^2-k^2\right)}{\left(p^2-m_{\rm dyn}^2\right)\left[\left(p+k\right)^2-m_{\rm dyn}^2\right]}\right]^{-1}\nn
&=&2\pi^2i\left[\left(4m_{\rm dyn}^2-k^2\right)\int_0^1 dx\ln\left(\frac{C_{\rm off}^2}{m_{\rm dyn}^2-x\left(1-x\right)k^2}\right)+\mbox{finite}\right]^{-1}~,
\eea
where $C_{\rm off}$ is again a UV cut off. The bound state wave function normalisation is then
\begin{align} \label{z}
Z&=\frac{1}{2\pi^2}\int_0^1 dx\ln\left(\frac{C_{\rm off}^2}{m_{\rm dyn}^2-x\left(1-x\right)k^2}\right)\\\nonumber
&=\frac{1}{2\pi^2}\ln\left(\frac{C_{\rm off}^2}{m_{\rm dyn}^2}\right)+{\cal O}\left(k^2\right)
\simeq-\frac{1}{2\pi^2}\ln\left(\omega^{2}\right)~,
\end{align}
which naturally inherits the multivalued structure of \eqref{musolution}.

An important comment is in order concerning the magnitude of $Z $ in (\ref{z}). In general, as we shall also discuss below, 
physically relevant situations, in the context of slow-roll inflationary phenomenology, require $Z > 1$. This may seem at first sight to contradict the usual unitarity bounds of $0 < Z <1$ imposed in field theory for fundamental fields. However,  here $Z$ refers to composite bound-state fields, for which such bounds are evaded~\cite{Bardeen,miranski,higash,alexvergou}.

We finally note that the mass prediction for the condensate, given by the pole $m_\sigma=2m_{\rm dyn}$ of the propagator \eqref{Gammap}, is not accurate and should be renormalised at the relevant infrared energy \cite{Bardeen}, 
in order to find a positive binding energy $B=2m_{\rm dyn}-m_\sigma>0$. 
Instead one should consider the mass obtained from the one-loop effective potential $V_{\text{eff}}(\sigma_c)$ found in ref.~\refcite{ahm}, and reviewed  in section \ref{sec:effpot}.

\section{Connection with Slow-Roll Inflation \label{sec:infl}}

Taking into account the results of the previous sections, the effective Lagrangian describing the gravitino bound state is 
\be
{\cal L}_{\rm eff}=\frac{Z\kappa^2}{2}\partial_\mu\sigma\partial^\mu\sigma-V_{\text{eff}}(\sigma)~,
\ee
where the rescaling $\sigma=\tilde\sigma/\kappa\sqrt{Z}$ leads to the canonically normalised Lagrangian
\be
\tilde{\cal L}_{\rm eff}=\frac{1}{2}\partial_\mu\tilde\sigma\partial^\mu\tilde\sigma-\tilde V_{\text{eff}}(\tilde\sigma)~,
\ee
and the coupling constants in the potential $\tilde V_{\text{eff}}$ are defined as
\be\label{Vn}
\tilde V_{\text{eff}}^{(n)}(0)\equiv\frac{V_{\text{eff}}^{(n)}(0)}{Z^{n/2}}~.
\ee
The latter normalisations ultimately yield the slow roll parameters 
\be\label{slowroll}
\epsilon=\frac{1}{Z}\frac{M_{\rm Pl}^2}{2}\left(\frac{V_{\text{eff}}'}{V_{\text{eff}}}\right)^2~,
~~~~\eta=\frac{1}{Z}M_{\rm Pl}^2\frac{V_{\text{eff}}''}{V_{\text{eff}}}~,
~~~~\xi=\frac{1}{Z^2}M_{\rm Pl}^4\frac{V_{\text{eff}}'V_{\text{eff}}'''}{V_{\text{eff}}^2}~.
\ee

As already mentioned, we assume that we can use the flat space-time estimates for the wave-function renormalisation obtained in the previous section, as expressing a correct order of magnitude estimate that is valid in the curved space-times during the inflationary period~\cite{emdyno}. We first notice, that in the broken phase, with the phenomenologically acceptable values of the gravitino mass $m_{\rm dyn}$ and supersymmetry breaking scales $\sqrt{f}$
(\ref{gravinoconf}), (\ref{fscaleconf}), the function $Z$ is of order one, which is consistent with the exit from the slow-roll inflationary phase.

To estimate the $Z$ near the origin of the potential (\ref{effpotsugra}), we use the expression (\ref{z}), but we replace the gravitino mass $m_{\rm dyn}$ by a transmutation mass scale $\tilde\mu$. 
\be\label{wfinfl}
Z \simeq-\frac{1}{2\pi^2}\ln\left(\omega_{\tilde \mu}^{2}\right)~, \quad \omega_{\tilde \mu} \equiv {\tilde \mu}/C_{\rm off}, \, 
\quad g\equiv\lambda_{\rm S} C_{\rm off}^2/2\pi^2~.
\ee  
It is now straightforward to see that only one branch of \eqref{musolution} (and equivalently \eqref{z}) is admissible for our purposes, with the $\omega$ variable replaced now by $\omega_{\tilde \mu}$, as defined in (\ref{wfinfl}). For the upper branch of figure \ref{fig:muplot} we have $\omega_{\tilde \mu} \to 1$ as $g\to1$ (we remind the reader that yielding $Z\to0$. 
This is difficult to reconcile with phenomenological values for our slow roll parameters \eqref{slowroll}.
In contrast, on the lower branch of figure \ref{fig:muplot} we have $\omega\to 0$ as $g\to1$. 
For appropriate values of $g$ this corresponds to the limit of large $Z \gg 1$ and small $\omega_{\tilde \mu}$, both of which are phenomenologically desirable~\footnote{As already pointed out, larger than one values of the wave-function renormalisation 
for the composite gravitino condensate fields do not contradict unitarity. A similar situation is encountered in composite Higgs symmetry breaking models in field theory~\cite{Bardeen}.}.

That large values of $Z \gg 1$ are necessarily linked to slow-roll hilltop inflation in this case is to be expected from the fact that the  effective potential (\ref{effpotsugra}) can be approximated near the origin (\emph{i.e}. for 
small field values of the condensate $\tilde\sigma \to 0$) as:
\be
V_{\rm eff} \simeq f^2 - (Z\kappa^2)^{-1} {\tilde  \sigma}^2 ~, \quad \tilde \sigma \to 0~, 
\ee
for a canonically normalised condensate field $\tilde \sigma$. 
To ensure that the slow-roll parameter $|\eta| < 1$ (\ref{slowroll}), then,
we must have
\be\label{estimates} 
Z \gg \frac{M_{\rm Pl}^4}{f^2} ~.
\ee
Since the (observed) running spectra index is of order $\eta_s \simeq 0.96$~\cite{Planck}, we must further impose that $|\eta| < 10^{-2}$. Phenomenologically realistic models of broken SUSY have $\sqrt{f} < 10^{16}$~GeV = $10^{-2}~M_{\rm Pl}$ (\emph{cf}. (\ref{fscaleconf})), hence we must have $Z \gg 10^{10}$, implying very small, practically vanishing, transmutation mass scales. 

\begin{figure}[h!!]
  \centering
    	\includegraphics[width=0.7\textwidth]{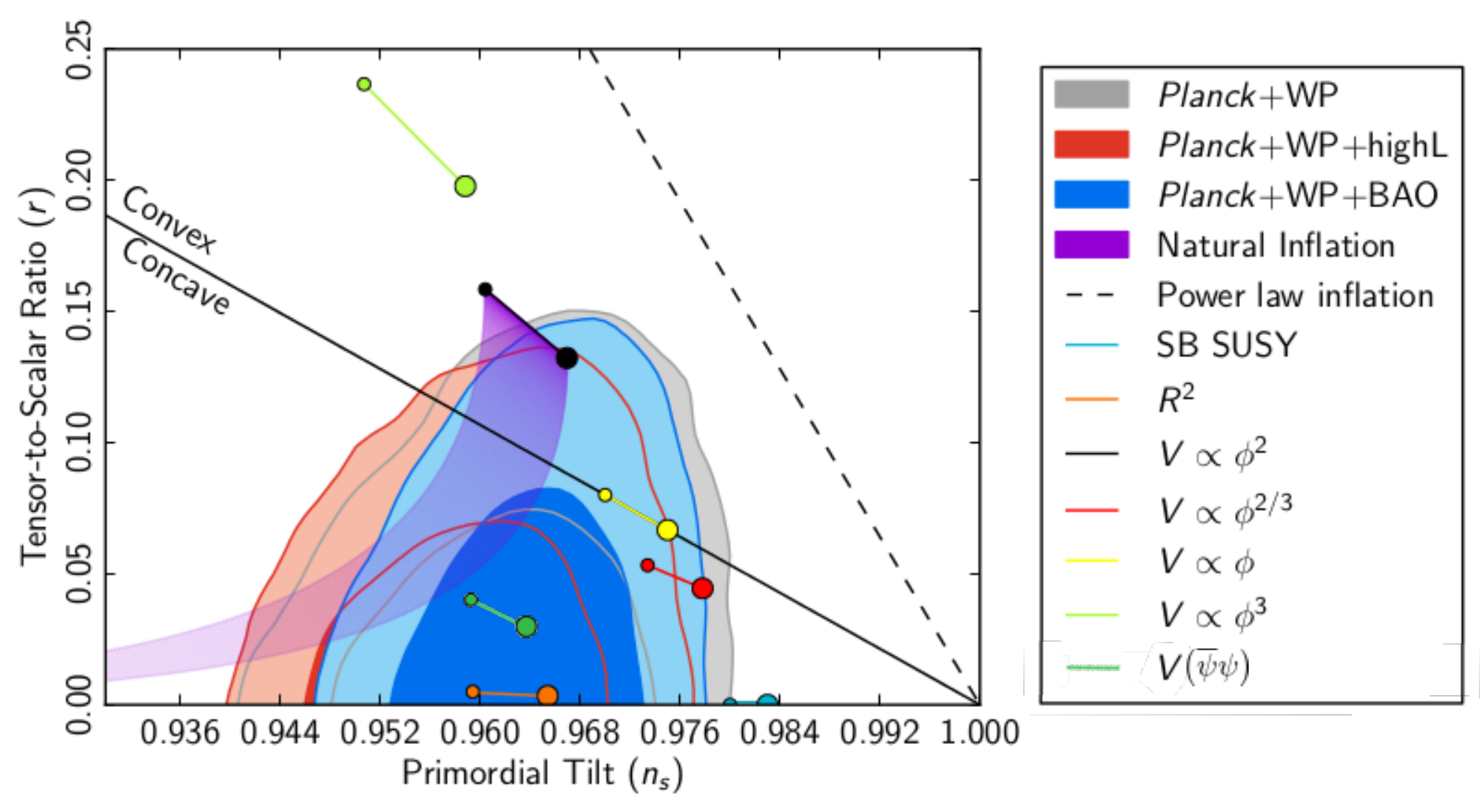}
    \caption{Planck data~\protect\cite{Planck} for $n_s$ and $r$,  as in fig.~\ref{fig:planck}, but with the gravitino-condensate hill-top inflation indicated explicitly (dark green). The latter model leads to higher $r$ than Starobinsky-type $R^2$ inflation (orange), although requires a very high value of the gravitino-condensate wave-function renormalisation, of order larger than $O(10^{16})$.}
    \label{fig:planckexclude}
\end{figure}

A typical case, compatible with the phenomenologically acceptable values
(\ref{fscaleconf}) and (\ref{gravinoconf}) is given in figure \ref{fig:planckexclude}, from which we observe that
agreement with Planck results is achieved for values of the wave function renormalisation of order
$Z \sim {\mathcal O }\left(10^{16}\right)$ 
for the phenomenologically relevant values of the couplings 
${\tilde \kappa }/\kappa \sim 10^3$. This corresponds to practically zero transmutation mass scales of $\tilde \mu \to 0$~\footnote{It may be interesting to notice that increasing ${\tilde \kappa }/\kappa$ higher still has the effect of scaling $V_{\rm eff}$ whilst leaving the shape of the potential qualitatively unchanged, allowing smaller and smaller values of $\sqrt{f}$ and $m_{\rm dyn}$.
Whilst this decrease in $f$ tends to naturally increase the slow-roll parameter $\eta$, by virtue of \eqref{estimates} this scenario may still be rendered compatible with slow-roll inflation if $Z$ is scaled accordingly to counteract this. 
As such, Planck compatible inflation as demonstrated in figure \ref{fig:planckexclude} can be achieved for any value of ${\tilde \kappa }/\kappa$. 
The Planck-compatible result is $(0.959, 0.04)\le \{n_s, r\} \le (0.964, 0.03)$ for 50 and 60 e-folds, respectively, corresponding to $\sqrt{f} \sim 5 \, e^{\langle \varphi \rangle} \times 10^{18}$ GeV. This is the case for any value of the (negative) dilaton vev $\langle \varphi \rangle$, however, as already mentioned, for realistic supersymmetry breaking phenomenology one should really fix $\sqrt{f}$ around or below the GUT scale.}.  

One way to interpret this result is the following. 
Near the origin of the potential one is in the unbroken phase, and hence the gravitino condensate has not yet fully formed, or rather is beginning to form, corresponding to a very small value of the gravitino mass. 
This small value grows in actual time, until the condensate sits in the minimum of the potential after rolling downhill, at which point the gravitino mass is stabilised at phenomenologically acceptable value, \emph{e.g}. of order the GUT scale. 
The duration of the whole process is that of the slow-roll inflation period, and exit from this phase occurs near the non-trivial minimum of the potential (\ref{effpotsugra}).

For our super-Higgs double well inflation it is difficult however to reconcile this model with the recent BICEP2 result of $r=0.16_{-0.05}^{+0.06}$ (after foreground subtraction)\cite{BICEP2}, 
since $V>0$ and $V''<0$ about the origin (and the wave function renormalisation $Z$ is necessarily positive), and thus $\eta<0$. 

Furthermore, although above we have presented hill-top inflationary models compatible with Planck results, associated with small field inflation at the origin of the potential (\ref{effpotsugra}), one may object to the huge value of the wave function renormalisation (\ref{estimates}) during the slow-roll inflationary phase~\footnote{One may be tempted to discuss, within the context of our minimal model, an alternative scenario, according to which global SUSY breaks at a transplanckian scale $\sqrt{f} \gg 1$ (in Planck units). In this case, the SUSY matter partners would completely decouple from the low-energy spectrum,
and hence there would be no experimental evidence for SUSY. On the other hand, local SUSY (SUGRA) would ensure inflation via the gravitino condensation mechanism described in this work, while the induced transplanckian dynamical mass for the gravitino, would remove any possibility of observing it as well.   From (\ref{estimates}) we can then conclude that slow-roll inflation could be achieved for natural values of the wave-function renormalisation $Z < {\mathcal O}(10)$, but in this case the stability of the electroweak vacuum would be delinked from any SUSY arguments. 
One could also try to relax the slow-roll assumption but this opens up a whole new game, where comparison with data may be complicated, 
and we do not consider it here.}.

There are however alternative scenarios of slow-roll inflation linked to this model which do not require such large $Z$, which we shall now come to discuss. 
These are associated with another type of inflation that may occur in the broken SUGRA phase, where, in contrast to the hill-top inflationary scenario discussed so far, the gravitino condensate field lies near its value that minimises the potential (\ref{effpotsugra}). 
In this scenario, the inflaton field is not the gravitino condensate, but it is linked to the scalar mode that parametrises a $R^2$-Starobinsky-like~\cite{staro} inflation that is associated with the effective gravitational action 
obtained after integrating out the massive gravitino-condensate degrees of freedom. 
This scenario 
was discussed in detail in ref.~\refcite{ahmstaro}, and we now proceed to review it briefly.  

\section{Starobinsky-type inflation in the broken SUGRA phase \label{sec:star}}

Starobinsky inflation is a model for obtaining a de Sitter (inflationary) cosmological solution to gravitational equations arising from a (four space-time-dimensional) action that includes higher curvature terms. 
Specifically, an action of the type in which the quadratic curvature corrections consist only of scalar curvature terms~\cite{staro}
\begin{eqnarray}\label{staroaction}
{\mathcal S} = \frac{1}{2 \, \kappa^2 } \, \int d^4 x \sqrt{-g}\,  \left(R  + \beta  \, R^2 \right) ~,~ 
\beta = \frac{8\, \pi}{3\, {\mathcal M}^2 }~,
\end{eqnarray}
where $\kappa^2=8\pi G$, and ${\rm G}=1/m_P^2$ is Newton's (gravitational) constant in four space-time dimensions, with $m_P$ the Planck mass, and ${\mathcal M}$ is a constant of mass dimension one, characteristic of the model. 

The important feature of this model is that inflationary dynamics are driven purely by the gravitational sector, through the $R^2$ terms, 
and that the scale of inflation is linked to ${\mathcal M}$. From a microscopic point of view, the scalar curvature-squared terms in (\ref{staroaction}) are viewed as the result of quantum fluctuations (at one-loop level)  of conformal (massless or high energy) matter fields of various spins, which have been integrated out in the relevant path integral in a curved background space-time~\cite{loop}. The quantum mechanics of this model, proceeding by means of tunnelling of the Universe from a state of ``nothing'' to the inflationary phase of ref.~\refcite{staro} has been discussed in detail in ref.~\refcite{vilenkin}.
The above considerations necessitate truncation to one-loop quantum order and to curvature-square (four-derivative) terms, which 
implies that there must be a region of validity for curvature invariants such that $\mathcal{O}\big(R^2/m_p^4\big) \ll 1$. 
This is of course a condition satisfied in phenomenologically realistic scenarios of inflation~\cite{Planck,encyclo}, for which the inflationary Hubble scale $H_I \leq  0.74 \times 10^{-5} \, m_P = {\mathcal O}(10^{15})~{\rm GeV}$ (the reader should recall that $R \propto H_I^2$ in the inflationary phase). 

\begin{figure}[h!!!]
\centering
		\includegraphics[width=0.5\textwidth]{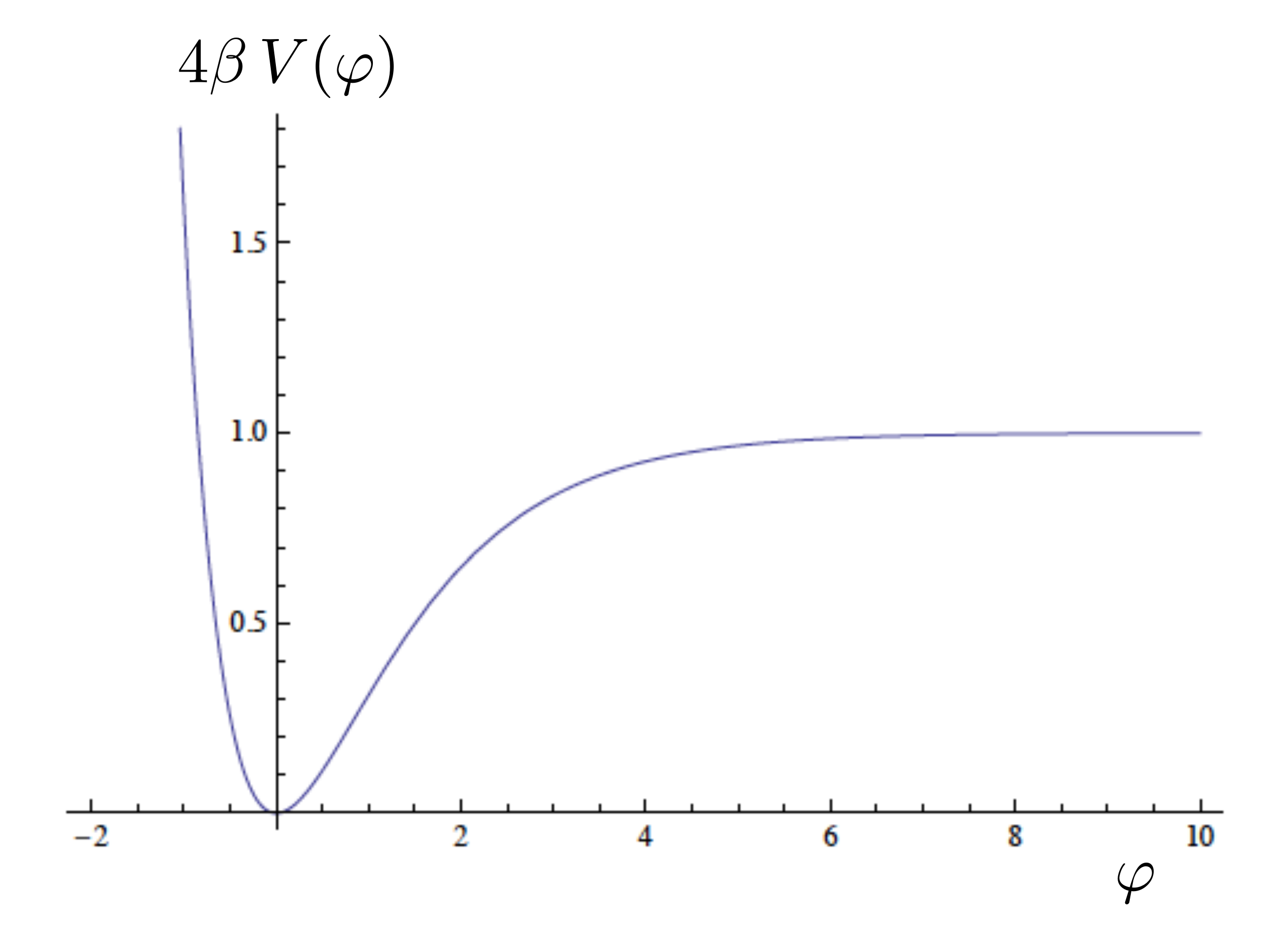}
		\caption{The effective potential (\ref{staropotent}) of the collective scalar field $\rho$ that describes the one-loop quantum fluctuations of matter fields, leading to the higher-order scalar curvature corrections in the Starobinski model for inflation (\ref{staroaction}). The potential is sufficiently flat to ensure slow-roll conditions for inflation are satisfied, in agreement with the Planck data, for appropriate values of the scale $1/\beta \propto {\mathcal M}^2$ (which sets the overall scale of inflation in the model).}
\label{fig:potstar}	
\end{figure}

Although the inflation in this model is not driven by rolling scalar fields, nevertheless the model (\ref{staroaction}) (and for that matter, any other model where the Einstein-Hilbert space-time Lagrangian density is replaced by an arbitrary function $f(R)$ of the scalar curvature) is conformally equivalent to that of an ordinary Einstein-gravity coupled to a scalar field with a potential that drives inflation~\cite{whitt}. 
To see this, one firstly linearises the $R^2$ terms in (\ref{staroaction}) by means of an auxiliary (Lagrange-multiplier) field $\tilde \alpha (x)$, before rescaling the metric by a conformal transformation and redefining the scalar field (so that the final theory acquires canonically-normalised Einstein and scalar-field terms):
\begin{eqnarray}\label{confmetric}
&&g_{\mu\nu} \rightarrow g^E_{\mu\nu} = \left(1 + 2 \, \beta \, {\tilde \alpha (x)} \right) \, g_{\mu\nu} ~, \quad
 \tilde \alpha \left(x\right) \to \rho (x) \equiv \sqrt{\frac{3}{2}} \, {\rm ln} \, \left(1 + 2\, \beta \, {\tilde \alpha \left(x\right)} \right)~.
\end{eqnarray}
These steps may be understood schematically via
\begin{align}\label{steps}
	&\int d^4 x \sqrt{-g}\,  \left( R  + \beta  \, R^2 \right) \\\nonumber
  	&\hookrightarrow\int d^4 x \sqrt{-g}\,  \left(  \left(1 + 2\, \beta \, \tilde \alpha \left(x\right) \right) \, R  -  \beta  \, {\tilde \alpha (x)}^2 \right)\\\nonumber
   &\hookrightarrow\int d^4 x \sqrt{-g^E}\,  \left(R^E +  g^{E\, \mu\, \nu} \, \partial_\mu \, \rho \, \partial_\nu \, \rho - V\right(\rho\left) \right)~,
\end{align}
where the arrows have the meaning that the corresponding actions appear in the appropriate path integrals, 
with the potential $V(\rho)$ given by:
\begin{eqnarray}\label{staropotent}
 V(\varphi ) = \frac{\left( 1 - e^{-\sqrt{\frac{2}{3}} \, \rho } \right)^2}{4\, \beta} \, 
  = \frac{3 {\mathcal M}^2 \, \Big( 1 - e^{-\sqrt{\frac{2}{3}} \, \rho } \Big)^2}{32\, \pi }  \,  ~.
\end{eqnarray}
The potential is plotted in fig.~\ref{fig:potstar}. 
We observe that it is sufficiently flat for large values of $\rho$ (compared to the Planck scale) to produce phenomenologically acceptable inflation, with the scalar field $\rho$ playing the role of the inflaton. 
In fact, the Starobinsky model fits the Planck data on inflation~\cite{Planck} well~\footnote{Quantum-gravity corrections in the original Starobinsky model \eqref{staroaction} have been considered recently in ref.~\refcite{copeland}, from the point of view of an exact renormalisation-group (RG) analysis~\cite{litim}. 
It was shown that the non-perturbative beta-functions for the `running' of Newton's `constant' G and the dimensionless $R^2$ coupling $\beta^{-1}$ in (\ref{staroaction}) imply an asymptotically safe ultraviolet (UV) fixed point 
for the former (that is, G($k \to \infty$) $ \to $ constant, for some 4-momentum cutoff scale $k$)), in the spirit of Weinberg~\cite{weinberg}, and an
attractive asymptotically-free ($\beta^{-1} (k \to \infty) \to 0$)
point for the latter. 
In this sense, the smallness of the $R^2$ coupling, required for agreement with
inflationary observables~\cite{Planck}, is naturally ensured by the presence of the asymptotically free UV fixed
point.}.

The agreement of the model of ref.~\refcite{staro} with the Planck data has triggered an enormous interest in the current literature
in revisiting the model from various points of view, such as its connection with no-scale supergravity~\cite{sugra_staro} and (super)conformal versions of supergravity and related areas~\cite{sugra_chaotic,sugrainfl}. 
In the latter works however the Starobinsky scalar field is fundamental, arising from the appropriate scalar component of some chiral superfield that appears in the superpotentials of the model. 
Although of great value, illuminating a connection between supergravity models and inflationary physics, and especially for explaining the low-scale of inflation compared to the Planck scale, it can be argued that these works contradict the original spirit of the Starobinsky model (\ref{staroaction}). 
Therein, higher curvature corrections are viewed as arising from quantum fluctuations of matter fields in a curved space-time background, such that inflation is driven by the pure gravity sector in the absence of fundamental scalars. 

In this section we consider an extension of the analysis of ref.~\refcite{ahm}, where the de Sitter parameter $\Lambda$ is perturbatively small compared to $m_P^2$, but not zero, so that truncation of the series to order $\Lambda^2$ suffices. 
This is in the spirit of the original Starobinsky model~\cite{staro}, with the role of matter fulfilled by the now-massive gravitino field.
Specifically, we are interested in the behaviour of the effective potential near the non-trivial minimum, where $\sigma_c $ is a non-zero constant. 
In our analysis, unlike Starobinsky's original work, we will keep the contributions from both graviton (spin-two) and gravitino quantum fluctuations.

We firstly note that the one-loop effective potential, obtained by integrating out gravitons and (massive) gravitino fields in the scalar channel (after appropriate euclideanisation), may be expressed as a power series in $\Lambda$: 
\begin{align}\label{effactionl2}
	\Gamma\simeq S_{\rm cl}-\frac{24\pi^2}{\Lambda^2 }\big(&\alpha^F_0+\alpha_0^B
	+ \left(\alpha^F_{1}+ \alpha^B_{1}\right)\Lambda
	+\left(\alpha^F_{2}+ \alpha^B_{2}\right)\Lambda^2+\dots\big)~,
\end{align} where $S_{\rm cl}$ denotes the classical action with tree-level cosmological constant $\Lambda_0$: 
	\begin{align}
		-\frac{1}{2\kappa^2}\int d^4 x \sqrt{g}\left(\widehat{R}-2\Lambda_0\right), \quad
		\Lambda_0\equiv\kappa^2\left(\sigma^2-f^2\right)~,
	\end{align} 
with $\widehat R$ denoting the fixed $S^4$ background we expand around ($\widehat R=4\Lambda$, Volume = $24\pi^2/\Lambda^2$), and the $\alpha$'s indicate the bosonic and fermionic quantum corrections at each order in $\Lambda$.

The leading order term in $\Lambda$ is then the effective action found in \cite{ahm} in the limit $\Lambda\to0$, 
	\begin{align}
		\Gamma_{\Lambda\to0}\simeq-\frac{24\pi^2}{\Lambda^2}\left(-\frac{\Lambda_0}{\kappa^2}+\alpha_0^F+\alpha_0^B\right)
		\equiv\frac{24\pi^2}{\Lambda^2}\frac{\Lambda_1}{\kappa^2},
	\end{align}
and the remaining quantum corrections then, proportional to $\Lambda$ and $\Lambda^2$ may be identified respectively with Einstein-Hilbert $R$-type and Starobinsky $R^2$-type terms in an effective action (\ref{effactionl3}) of the form
\begin{align}\label{effactionl3}
\Gamma\simeq&-\frac{1}{2\kappa^2} \int d^4 x \sqrt{g} \left(\left(\widehat R-2\Lambda_1\right)  +\alpha_1 \, \widehat R+ \alpha_2 \, \widehat R^2\right)~,
\end{align}
where we have combined terms of order $\Lambda^2$ into curvature scalar square terms. For general backgrounds such terms 
would correspond to invariants of the form ${\widehat R}_{\mu\nu\rho\sigma} \, {\widehat R}^{\mu\nu\rho\sigma} $, ${\widehat R}_{\mu\nu} \, {\widehat R}^{\mu\nu}$ and ${\widehat R}^2$, which for a de Sitter background all combine to yield ${\widehat R}^2$ terms. 
The coefficients $\alpha_1$  and $\alpha_2$ absorb the non-polynomial (logarithmic) in $\Lambda$ contributions, so that we may then identify \eqref{effactionl3} with \eqref{effactionl2} via 
	\begin{align}\label{alpha}
		\alpha_1=\frac{\kappa^2}{2}\left(\alpha^F_1+\alpha^B_1\right)~,\quad
		\alpha_2=\frac{\kappa^2}{8}\left(\alpha^F_2+\alpha^B_2\right)~.
	\end{align}

To identify the conditions for phenomenologically acceptable Starobinsky inflation around the non-trivial minima of the broken SUGRA phase 
of our model, we impose first the cancellation of the ``classical'' Einstein-Hilbert space term $\widehat R $ by the ``cosmological constant'' term $\Lambda_1$, i.e. that $\widehat R = 4 \, \Lambda = 2\, \Lambda_1 $.
This condition should be understood as a necessary one characterising our background in order to produce phenomenologically-acceptable 
Starobinsky inflation in the broken SUGRA phase following the first inflationary stage, as discussed in ref.~\refcite{emdyno}. 
This may naturally be understood as a generalisation of the relation $\widehat R=2\Lambda_1=0$, imposed in ref.~\refcite{ahm} as a self-consistency condition for the dynamical generation of a gravitino mass.

The effective Newton's constant in  (\ref{effactionl3}) is then $\kappa_{\rm eff}^2=\kappa^2/\alpha_1$, and from this, we can express the effective Starobinsky scale (\ref{staroaction}) in terms of $\kappa_{\rm eff}$ as $\beta_{\rm eff} \equiv  \alpha_2/\alpha_1$.
This condition thus makes a direct link between the action (\ref{effactionl2}) with a Starobinsky type action (\ref{staroaction}).
Comparing with (\ref{staroaction}), we can then identify the Starobinsky inflationary scale in this case as
\begin{equation}\label{staroours}
{\mathcal M} = \sqrt{\frac{8 \pi}{3} \, \frac{\alpha_1}{\alpha_2} }~.
\end{equation}

We may then determine the coefficients $\alpha_1$ and $\alpha_2$ in order to evaluate the scale $1/\beta$ of the effective Starobinsky potential given in fig.~\ref{fig:potstar} in this case, and thus the scale of the second inflationary phase. 
To this end, we use the results of ref.~\refcite{ahm}, derived via an asymptotic expansion as explained in the appendix therein, to obtain the following forms for the coefficients 		
\begin{align}\label{aif}
		\alpha^F_1&=\frac{\left(25491-5 \sqrt{27076337}\right)}{25016} \tilde\kappa^2 \sigma_c ^2 \log \left(\frac{\Lambda}{\mu^2}\right) 
		+\frac{\left(3 \sqrt{65028102}-18700\right) }{81397}\tilde\kappa^2 \sigma_c ^2\\\nonumber
		&+\frac{\left(\sqrt{100304662585}-247787\right) }{945888}\tilde\kappa^2 \sigma_c ^2 \log \left(\frac{\tilde\kappa^2\sigma_c^2}{\mu^2} \right)~,  \nonumber
	\end{align}
	\begin{align}
		\alpha^F_{2}&=\frac{\left(6 \sqrt{5018206}-12882\right)}{38914}\log \left(\frac{\tilde\kappa^2\sigma_c^2}{\mu^2}\right)
		+\frac{\left(50249-\sqrt{2590498021}\right) }{22066}\log \left(\frac{\Lambda}{\mu^2}\right)\\\nonumber
		&+\frac{\sqrt{10592733}-1377}{65388}~,
	\end{align}
and 
	\begin{align}\label{aib}
		\alpha^B_1&=
		\frac{\sqrt{356979979}-17707}{64839}\Lambda_0 \log \left(\frac{\Lambda }{3 \mu ^2}\right) 		
		+\frac{\left(\sqrt{2812791101}-52583\right) }{9244}\Lambda_0 \log \left(-\frac{3 \Lambda_0}{\mu ^2}\right)\\\nonumber
		&-\frac{\left(\sqrt{1416210349}-27907\right)\left(1+\log\left(2\right)\right)}{198570} \Lambda_0~, \nonumber
	\end{align}
	\begin{align}
		\alpha^B_{2}&=-\frac{\left(\sqrt{220573721}-19811\right) }{232300}\log \left(\frac{\Lambda }{3 \mu ^2}\right)
		+\frac{\left(10 \sqrt{12614479}-36763\right) }{86027}\log \left(-\frac{6 \Lambda_0}{\mu ^2}\right)	\\\nonumber
		&+\frac{2731-\sqrt{1392978}}{76777}~,
	\end{align}
where ${\tilde \kappa} = e^{-\langle \varphi  \rangle } \, \kappa$ is the conformally-rescaled gravitational constant in the model of ref.~\refcite{confsugra}, defined previously via \eqref{tildcoupl}. In the case of standard ${\mathcal N}=1$ SUGRA, $\langle \varphi \rangle = 0$.  

We note at this stage that the spin-two parts, arising from integrating out graviton quantum fluctuations, are not dominant in the conformal case~\cite{ahm}, provided ${\tilde \kappa}/\kappa \ge {\mathcal O}(10^3)$, which leads~\cite{emdyno} to the agreement of the first  inflationary phase of the model with the Planck data~\cite{Planck}.  
However, if the first phase is succeeded by a Starobinsky phase, it is the latter only that needs to be checked against the data. 

To this end we search numerically for points in the parameter space such that; the effective equations
	\begin{align}
		 \frac{\partial\Gamma}{\partial\Lambda}=0~, \quad
		 \frac{\partial\Gamma}{\partial\sigma}=0~,
	\end{align}
are satisfied, $\Lambda$ is small and positive ($0<\Lambda<10^{-5}M^2_{\rm Pl}$, to ensure the validity of our expansion in $\Lambda$) and $10^{-6}<\mathcal{M}/M_{\rm Pl}<10^{-4}$, to match with known phenomenology of  Starobinsky inflation \cite{Planck}. 

For $\tilde \kappa=\kappa$ (i.e. for non-conformal supergravity), we were unable to find any solutions satisfying these constraints. 
This of course may not be surprising, given the previously demonstrated non-phenomenological suitability of this simple model~\cite{ahm}. 

\begin{figure}[h!!!]
	\centering
		\includegraphics[width=0.48\textwidth]{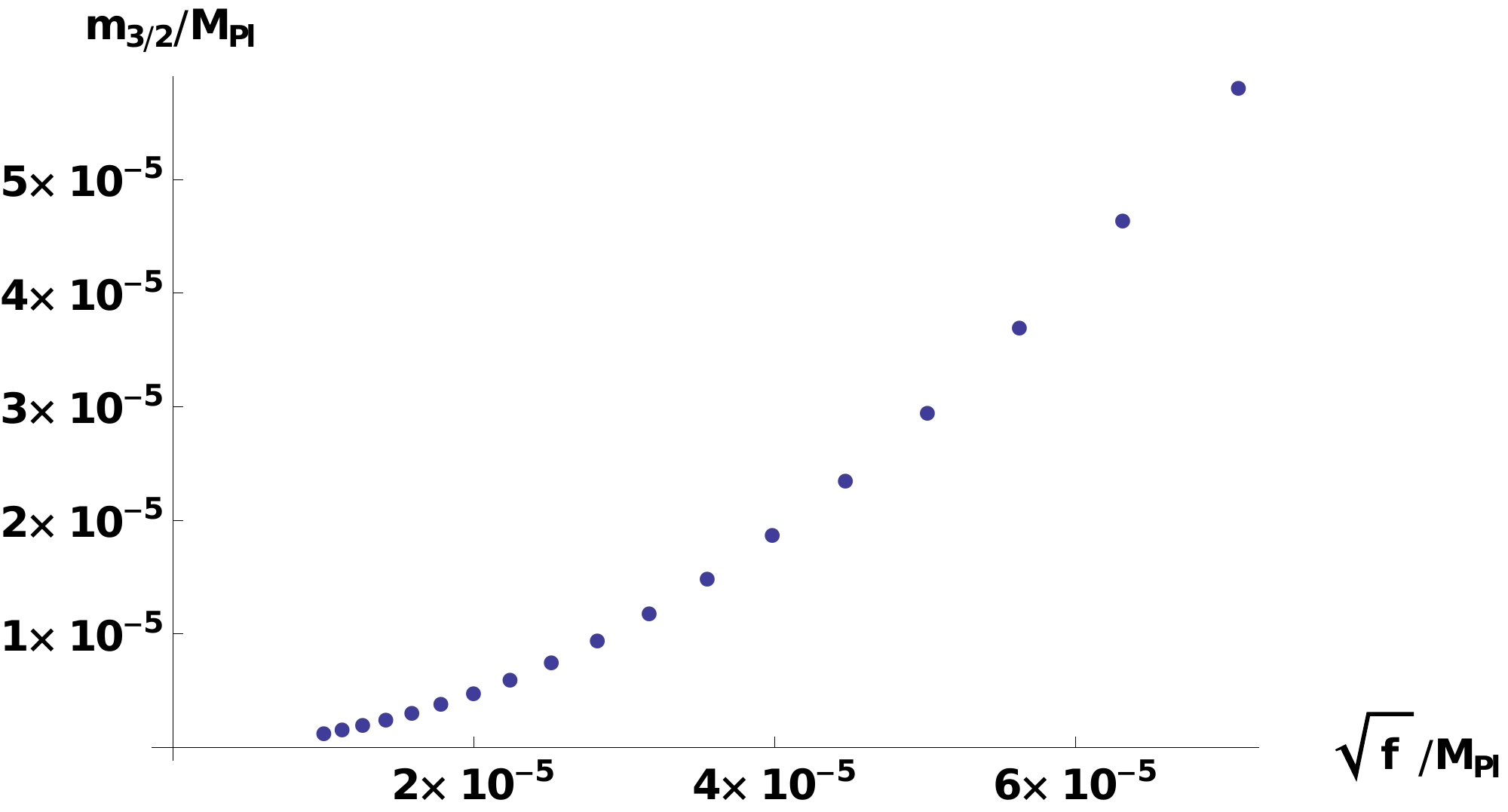} \hfill
		\includegraphics[width=0.48\textwidth]{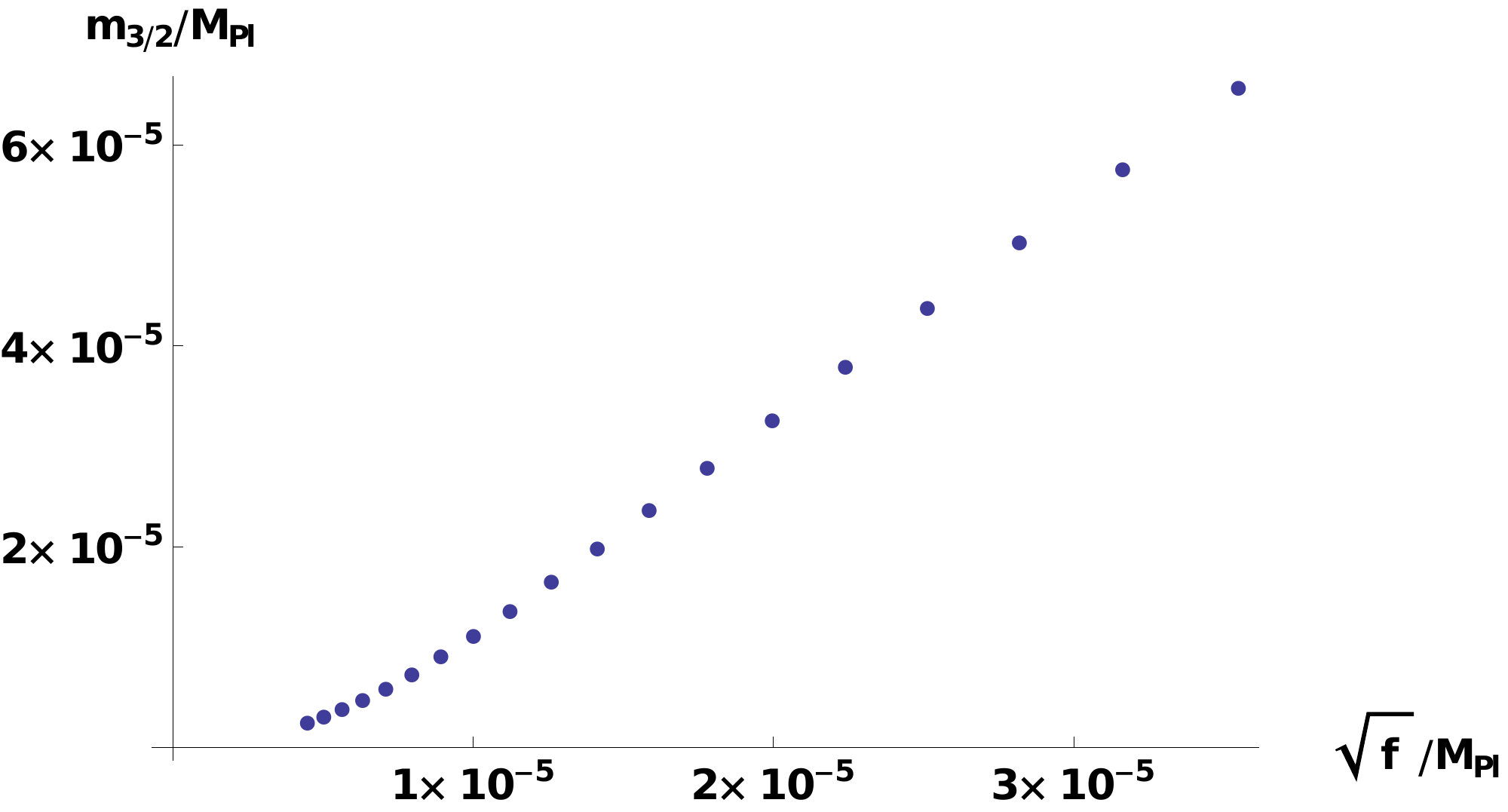}		
		\caption{\emph{Left panel: }Results for $\tilde \kappa=10^3\kappa$. \emph{Right panel:} Results for $\tilde \kappa=10^4\kappa$.}
		\label{fig:2a}
		\end{figure}

If we consider $\tilde \kappa>>\kappa$ however, we find that we are able to satisfy the above constraints for a range of values. 
We present this via the two representative cases below, indicated in fig.~\ref{fig:2a}, where
$\sqrt{f}$ is the scale of global supersymmetry breaking, and we have set the normalisation scale via $\kappa\mu=\sqrt{8\pi}$.\\
Every point in the graphs of the figures is selected to make the Starobinsky scale of order ${\mathcal M} \sim 10^{-5} \, M_{\rm Pl} $, so as to be able to achieve phenomenologically acceptable inflation in the massive gravitino phase, consistent with the Planck-satellite data~\cite{Planck}. 

Exit from the inflationary phase is a complicated issue which we shall not discuss here, aside from the observation that it can be achieved by
coherent oscillations of the gravitino condensate field around its minima, or tunnelling processes \`a  la Vilenkin~\cite{vilenkin}. 
This is still an open issue, which may be addressed via construction of more detailed supersymmetric models, including coupling of the matter sector to gravity. 

\section{Conclusions \label{sec:concl}}

Inflation (of hilltop type) via gravitino condensate is a well motivated scenario to both inflate the early universe and break local supersymmetry, although one which, whilst supported by the Planck satellite results \cite{Planck}, is not at present favoured by recent observations of a large tensor to scalar ratio by the BICEP2 collaboration \cite{BICEP2}.

In light of this tension between the BICEP2 and Planck results, we have addressed in this paper the requisite technical steps to connect this scenario with precision inflationary phenomenology; specifically the calculation of the condensate wave function renormalisation, and a Schwinger-Dyson based resolution of the Fierz ambiguity inherent to this mean field theory approach. Once more data become available, hopefully reducing or eliminating this tension, the viability of this model may then be confronted fully by observations.

Nevertheless, given that agreement of this hilltop model for inflation with slow-roll scenarios requires (for the case of phenomenologically realistic supersymmetry breaking scales at or below the grand unification scale) very large values of the gravitino-condensate wave function renormalisation, that may be considered as unnaturally high, we have considered alternative scenarios of inflation of Starobinsky-type. 
The latter may occur in the broken SUGRA phase, as a result of integrating out the massive gravitino fields in the gravitational effective action. 

This second scenario, which is also in agreement with the Planck data,  does not suffer from any unnaturally large or small parameters and in this sense it may be preferred by some to the hill-top inflation. However, 
it leads to a much more suppressed tensor-to-scalar ratio than the hill-top inflationary scenario, and in this sense 
is in stark disagreement with the BICEP2 results, although the latter have still to be confirmed by Planck and other future experiments.

\section*{Acknowledgements}

We thank  S.~Basilakos and E.~Plionis, editors of the special volume of IJMPD on \emph{Inflation}, for their kind invitation to present a review of our work. The work of N.H. is supported by a KCL GTA studentship, while that of N.E.M. is supported in part by the London Centre for Terauniverse Studies (LCTS), using funding from the European Research Council via the Advanced Investigator Grant 267352 and by STFC (UK) under the research grant ST/J002798/1.

\section*{Appendix A: Fierz identities}

To assist in simplifying our fermion bilinears we may leverage some useful Fierz identities.
Firstly note that there exists an orthogonal basis for the Clifford algebra via antisymmetrised products of gamma matrices, i.e. for $D=4$, following the conventions of ref.~\refcite{Freedman}
	\begin{align}
		\left\{\Gamma^A=\mathbb{1},\gamma^{\mu_1},\gamma^{\mu_1\mu_2},\gamma^{\mu_1\mu_2\mu_3},\gamma^{\mu_1\mu_2\mu_3\mu_4}\right\}, 
		\quad \gamma^{\mu_1\dots\mu_n}\equiv\gamma^{[\mu_1}\dots\gamma^{\mu_n]}\, ,
	\end{align}
where antisymmetrisation is always with unit weight.
We may then construct the (index-reversed) dual basis 
	\begin{align}
		\left\{\Gamma_A=\mathbb{1},\gamma_{\mu_1},\gamma_{\mu_2\mu_1},\gamma_{\mu_3\mu_2\mu_1},\gamma_{\mu_4\mu_3\mu_2\mu_1}\right\}\, , 
	\end{align}
such that the trace orthogonality condition is $\Tr\left(\Gamma^A\Gamma_B\right)=4\delta^A{}_B$ is always satisfied.

Any matrix $M$ may be then expanded in this basis
\begin{align}	
	M=\sum_A m_A \Gamma^A\, ,
\end{align}
where the coefficients $m_A$ can be identified via
\begin{align}	
	\Tr\left(\Gamma_A M\right)
	=\Tr\left(\Gamma_A m_B \Gamma^B\right)
	=4 m_B \delta^B{}_A
	=4 m_A\, .
\end{align}
To make use of this, we may firstly write (suppressing Lorentz indices)
\begin{align}
	\left(\overline\lambda_1M\lambda_2\right)\left(\overline\lambda_3N\lambda_4\right)
	=\overline\lambda_{1\alpha}\lambda_2{}^\beta\overline\lambda_{3\gamma}\lambda_4{}^\delta M^\alpha{}_\beta N^\gamma{}_\delta\, ,
\end{align}
for some anticommuting spinors $\lambda_i$, and then identify $M^\alpha{}_\beta N^\gamma{}_\delta$ as a matrix $P_\beta{}^\gamma\left(\alpha,\delta\right)$ for fixed $\left\{\alpha, \delta\right\}$, giving  
\begin{align}
	P_\beta{}^\gamma\left(\alpha,\delta\right)
	&=\frac{1}{4}\sum_A\Tr\left(p^\varphi{}_\epsilon\left(\Gamma_A\right)^\epsilon{}_\rho\right)\left(\Gamma^A\right)_\beta{}^\gamma
	=\frac{1}{4}\sum_Ap^\rho{}_\epsilon\left(\Gamma_A\right)^\epsilon{}_\rho\left(\Gamma^A\right)_\beta{}^\gamma\\
	&=\frac{1}{4}\sum_A\left(M^\alpha{}_\epsilon N^\rho{}_\delta\left(\Gamma^A\right)^\epsilon{}_\rho\right)\left(\Gamma^A\right)_\beta{}^\gamma
	=\frac{1}{4}\sum_A\left(M\Gamma_A N\right)^\alpha{}_\delta\left(\Gamma^A\right)_\beta{}^\gamma \, .
\end{align}
This then yields the standard expansion of products of bilinears (noting a minus sign from anticommutativity of $\lambda_i$)
	\begin{align}
		&\left(\overline\lambda_1 M\lambda_2\right)\left(\overline\lambda_3 N\lambda_4\right)
		=-\frac{1}{4}\sum_A\left(\overline\lambda_1 M\Gamma_A N\lambda_4\right)\left(\overline\lambda_3\Gamma^A\lambda_2\right)\\\nonumber
		&=-\frac{1}{4}\sum_n\frac{1}{n!}\left(\overline\lambda_1M\gamma_{\mu_1\dots \mu_n}N\lambda_4\right)
		\left(\overline\lambda_3 \gamma^{\mu_n\dots \mu_1} \lambda_2\right)\, ,
	\end{align}
where the factor of $1/n!$ is introduced to avoid overcounting of the same $\gamma_{\mu_1\dots \mu_n}$ matrix $n!$ times.

In the case at hand, significant simplifications are possible since we only have one spinor; the gravitino.
Indeed, we may note that (no longer suppressing Lorentz indices)
\begin{align}
	\overline\lambda_{1\alpha} \gamma_{\mu_1}\gamma_{\mu_2}\dots\gamma_{\mu_n}\lambda_{2\beta}=
	\left(-1\right)^n\overline\lambda_{2\beta} \gamma_{\mu_n}\gamma_{\mu_{n-1}}\dots\gamma_{\mu_1}\lambda_{1\alpha}\, ,
\end{align}
which implies that
\begin{align}
	 \overline\lambda_{1\alpha}\gamma^{\mu_1}\lambda_{1}{}^\alpha
	 =\overline\lambda_{1\alpha}\gamma^{\mu_1\mu_2}\lambda_{1}{}^{\alpha}
	 =0\, ,
\end{align}
and we need only consider expansion in a subset of our basis elements.
Furthermore, we may note a trilinear identity from the Appendix of ref.~\refcite{Nieuwenhuizen}
\begin{align}\label{trilinear}
	\left(\overline\lambda_{1\mu}\lambda_1^\mu\right)\lambda_\alpha
	=-\left(\overline\lambda_{1\mu}\gamma^5\lambda_1^\mu\right)\left(\gamma^5\lambda_1\right)_\alpha
	=\frac{1}{4}\left(\overline\lambda_{1\mu}\gamma^5\gamma^\nu\lambda_1^\mu\right)\left(\gamma^5\gamma_\nu\lambda_1\right)_\alpha\,
\end{align}	
which, if we left multiply with $\overline\lambda_1^\alpha$, allows remaining basis elements, re-expressed via the useful identities 
\begin{align}
	\gamma_{\mu_1\mu_2\mu_3}=i\epsilon_{\mu_1\mu_2\mu_3\mu_4}\gamma^{\mu_4}\gamma^5, \quad
	\gamma_{\mu_1\mu_2\mu_3\mu_4}=-i\epsilon_{\mu_1\mu_2\mu_3\mu_4}\gamma^5\, ,
\end{align}
to be simplified further.

For our quantity of interest (noting the permutation of the first bilinear relative to \eqref{torsion})
\begin{align} 
	\mathcal{L}_{\rm torsion}=\frac{1}{16}\left(\left(\overline\psi^\nu\gamma^\mu\psi^\rho\right)\left(\overline\psi_\rho\gamma_\mu\psi_\nu+2\overline\psi_\rho\gamma_\nu\psi_\mu\right)\right)\times 2\kappa^2\,,
\end{align}
we may compute (using that $\gamma\cdot\psi=0$)
\begin{align}\nonumber
	&\left(\overline\psi^\nu\gamma^\mu\psi^\rho\right)\left(\overline\psi_\rho\gamma_\mu\psi_\nu\right)\\\nonumber
	&=-\left(\overline\psi^\nu\psi_\nu\right)\left(\overline\psi_\rho\psi^\rho\right)
	-\frac{1}{4}\left(\overline\psi^\nu\gamma^\mu\gamma^5\gamma^\alpha\gamma_\mu\psi_\nu\right)\left(\overline\psi_\rho\gamma^5\gamma_\alpha\psi^\rho\right)
	+\frac{1}{4}\left(\overline\psi^\nu\gamma^\mu\gamma^5\gamma_\mu\psi_\nu\right)\left(\overline\psi_\rho\gamma^5\psi^\rho\right)\\
	&=-\left(\overline\psi^\nu\psi_\nu\right)\left(\overline\psi_\rho\psi^\rho\right)
	-\frac{1}{2}\left(\overline\psi^\nu\gamma^5\gamma^\alpha\psi_\nu\right)\left(\overline\psi_\rho\gamma^5\gamma_\alpha\psi^\rho\right)
	-\left(\overline\psi^\nu\gamma^5\psi_\nu\right)\left(\overline\psi_\rho\gamma^5\psi^\rho\right)\,,\\
	&\left(\overline\psi^\nu\gamma^\mu\psi^\rho\right)\left(\overline\psi_\rho\gamma_\nu\psi_\mu\right)\\\nonumber
	&=-\frac{1}{2}\left(\overline\psi^\nu\psi_\nu\right)\left(\overline\psi_\rho\psi^\rho\right)
	-\frac{1}{4}\left(\overline\psi^\nu\gamma^\mu\gamma^5\gamma^\alpha\gamma_\nu\psi_\mu\right)\left(\overline\psi_\rho\gamma^5\gamma_\alpha\psi^\rho\right)
	+\frac{1}{4}\left(\overline\psi^\nu\gamma^\mu\gamma^5\gamma_\nu\psi_\mu\right)\left(\overline\psi_\rho\gamma^5\psi^\rho\right)\nonumber\\
	&=-\frac{1}{2}\left(\overline\psi^\nu\psi_\nu\right)\left(\overline\psi_\rho\psi^\rho\right)
	-\frac{1}{2}\left(\overline\psi^\nu\gamma^5\gamma^\alpha\psi_\nu\right)\left(\overline\psi_\rho\gamma^5\gamma_\alpha\psi^\rho\right)
	-\frac{1}{2}\left(\overline\psi^\nu\gamma^5\psi_\nu\right)\left(\overline\psi_\rho\gamma^5\psi^\rho\right)\, .
\end{align}

Simplifying via \eqref{trilinear} (noting in particular that the first and last terms in each line then cancel), we may then write 
\begin{align}
		\mathcal{L}_{\rm torsion}
		=-\frac{3}{8}\left(\overline\psi^\rho\psi_\rho\right)^2\times 2\kappa^2~,
\end{align}
which we made use in the text, cf. (\ref{couplings}).

  \end{document}